%% file: main_GLOBECOM_2019_-_Paper_1_-_CR.tex
\documentclass[conference]{IEEEtran}
\usepackage[utf8]{inputenc}
\usepackage[english]{babel}
\usepackage{algorithm}
\usepackage{algorithmicx}
\usepackage{algpseudocode}
\usepackage{amssymb}
\usepackage{array}
\usepackage{arydshln}
\usepackage{blkarray, bigstrut}
\usepackage{bm}
\usepackage{booktabs}
\usepackage{braket}
\usepackage{cite}
\usepackage{cuted}
\usepackage{enumitem}
\usepackage{graphicx}
\usepackage{gensymb}
\usepackage[mathscr]{euscript}
\usepackage[switch]{lineno}
\usepackage{mathtools}
\usepackage{multirow}
\usepackage{multicol}
\usepackage{mathtools}
\usepackage{xfrac}
\usepackage{framed} 
\usepackage[framed]{ntheorem}
\usepackage{pgfplots}
\usepgfplotslibrary{polar}
\usepackage{qtree}
\usepackage{adjustbox}
\usepackage{rotating}
\pgfplotsset{compat = 1.13}
\usepackage{stfloats}
\usepackage{subcaption}
\usepackage{url}
\usepackage{tikz}
\usepackage{tkz-berge}
\usepackage{fixltx2e}
\usepackage[framemethod=TikZ]{mdframed}
\usetikzlibrary{patterns}
\usetikzlibrary{spy}
\usetikzlibrary{shapes, backgrounds,calc, snakes}
\usetikzlibrary{decorations.pathreplacing}
\usetikzlibrary{matrix}
\usepackage{fancybox}
\usepackage{caption}
\usepackage{amsmath}

\tikzstyle{vertex} = [circle, draw, inner sep = 0pt, minimum size = 10pt]

\newframedtheorem{frm-prop}{Property}
\usepgfplotslibrary{
  units, 
  groupplots
}
\pgfplotsset{compat=1.11,
    /pgfplots/ybar legend/.style={
    /pgfplots/legend image code/.code={%
       \draw[##1,/tikz/.cd,yshift=-0.25em]
        (0cm,0cm) rectangle (5pt, 8pt);},
   },
}

\definecolor{bblue}{rgb}{0.12392, 0.0490, 0.9588}
\definecolor{sskyblue}{rgb}{0.1529, 0.5882, 0.9216}
\definecolor{ggreen}{rgb}{0.7098, 0.95, 0.40781}
\definecolor{yyellow}{rgb}{0.9765, 0.9804, 0.0784}

\definecolor{color0}{HTML}{FF0147}
\definecolor{color1}{HTML}{F400DC}
\definecolor{color2}{HTML}{BA0DFF}
\definecolor{color3}{HTML}{5700E8}
\definecolor{color4}{HTML}{0B03FF}
\definecolor{color5}{HTML}{0957F4}
\definecolor{color6}{HTML}{03B3FF}
\definecolor{color7}{HTML}{08E8DA}
\definecolor{color8}{HTML}{07FF8E}
\definecolor{color9}{HTML}{51FF0A}

\definecolor{p1}{rgb}{1, 0.0667, 0}
\definecolor{p2}{rgb}{1, 0.24, 0}
\definecolor{p3}{rgb}{1, 0.349, 0}
\definecolor{p4}{rgb}{1, 0.490, 0}
\definecolor{p5}{rgb}{1, 0.631, 0}
\definecolor{p6}{rgb}{1, 0.792, 0}
\definecolor{p7}{rgb}{1, 0.933, 0}

\definecolor{ccolor0}{HTML}{FF007D}
\definecolor{ccolor1}{HTML}{760CE8}
\definecolor{ccolor2}{HTML}{0A55FF}
\definecolor{ccolor3}{HTML}{0DB6F4}
\definecolor{ccolor4}{HTML}{00FF76}
\definecolor{ccolor5}{HTML}{6FE80C}
\definecolor{ccolor6}{HTML}{FFDE0A}
\definecolor{ccolor7}{HTML}{FF990A}

\definecolor{bcolor0}{HTML}{fff7fb}
\definecolor{bcolor1}{HTML}{ece7f2}
\definecolor{bcolor2}{HTML}{d0d1e6}
\definecolor{bcolor3}{HTML}{a6bddb}
\definecolor{bcolor4}{HTML}{74a9cf}
\definecolor{bcolor5}{HTML}{3690c0}
\definecolor{bcolor6}{HTML}{0570b0}
\definecolor{bcolor7}{HTML}{045a8d}
\definecolor{bcolor8}{HTML}{023858}

\definecolor{gcolor0}{HTML}{c8c8c8}
\definecolor{gcolor1}{HTML}{b5b5b5}
\definecolor{gcolor2}{HTML}{a3a3a3}
\definecolor{gcolor3}{HTML}{919191}
\definecolor{gcolor4}{HTML}{7e7e7e}
\definecolor{gcolor5}{HTML}{6c6c6c}
\definecolor{gcolor6}{HTML}{595959}
\definecolor{gcolor7}{HTML}{474747}

\definecolor{q0}{HTML}{4d4d4d}
\definecolor{q1}{HTML}{fef0d9}
\definecolor{q2}{HTML}{fdd49e}
\definecolor{q3}{HTML}{fdbb84}
\definecolor{q4}{HTML}{fc8d59}
\definecolor{q5}{HTML}{ef6548}
\definecolor{q6}{HTML}{d7301f}
\definecolor{q7}{HTML}{990000}

\definecolor{p0}{HTML}{b2182b}
\definecolor{p1}{HTML}{d6604d}
\definecolor{p2}{HTML}{f4a582}
\definecolor{p3}{HTML}{fddbc7}
\definecolor{p4}{HTML}{e0e0e0}
\definecolor{p5}{HTML}{bababa}
\definecolor{p6}{HTML}{878787}
\definecolor{p7}{HTML}{4d4d4d}

\definecolor{empcolor}{HTML}{a1382d}

\usepackage{todonotes}

\newcommand{\fref}[1]{Fig.~\ref{#1}}

\newcommand{\sref}[1]{Section~\ref{#1}}

\DeclareMathSizes{10pt}{8pt}{6pt}{5pt}
\newcommand{\CLASSINPUTbottomtextmargin}{0.8in}

\makeatletter
\newcommand\fs@betterruled{%
  \def\@fs@cfont{\bfseries}\let\@fs@capt\floatc@ruled
  \def\@fs@pre{\vspace*{8pt}\hrule height.8pt depth0pt \kern2pt}%
  \def\@fs@post{\kern2pt\hrule\relax}%
  \def\@fs@mid{\kern2pt\hrule\kern2pt}%
  \let\@fs@iftopcapt\iftrue}
\floatstyle{betterruled}
\restylefloat{algorithm}
\makeatother

\begin{document}
\title{Hybrid Precoding for Multi-Group Multicasting \\ in mmWave Systems}

\author{
\IEEEauthorblockN{Luis F. Abanto-Leon, Matthias Hollick, and Gek Hong (Allyson) Sim} 
\IEEEauthorblockA{Secure Mobile Networking (SEEMOO) Lab, Technische Universit\"{a}t Darmstadt, Germany}
\{labanto, mhollick, asim\}@seemoo.tu-darmstadt.de
}

\thanks{M. Shell was with the Department
of Electrical and Computer Engineering, Georgia Institute of Technology, Atlanta,
GA, 30332 USA e-mail: (see http://www.michaelshell.org/contact.html).}
\thanks{J. Doe and J. Doe are with Anonymous University.}
\thanks{Manuscript received April 19, 2005; revised August 26, 2015.}

\markboth{Journal of \LaTeX\ Class Files,~Vol.~14, No.~8, August~2015}%
{Shell \MakeLowercase{\textit{et al.}}: Bare Demo of IEEEtran.cls for IEEE Communications Society Journals}

\maketitle

\input{abstract}

\begin{IEEEkeywords}
hybrid precoding, millimeter-wave, multicast, semidefinite relaxation, alternating optimization.
\end{IEEEkeywords}

\IEEEpeerreviewmaketitle

\input{introduction}
\input{system_model}

\input{result}

\input{discussion}
\input{conclusion}
\input{acknowledgment}

\bibliographystyle{IEEEtran}
\bibliography{ref}

\end{document}

%% file: abstract.tex
\begin{abstract}
Multicast beamforming is known to improve spectral efficiency. However, its benefits and challenges for hybrid precoders design in millimeter-wave (mmWave) systems remain understudied. To this end, this paper investigates the first joint design of hybrid transmit precoders (with an arbitrary number of finite-resolution phase shifts) and receive combiners for mmWave multi-group multicasting. Our proposed design leverages semidefinite relaxation (SDR), alternating optimization and Cholesky matrix factorization to sequentially optimize the digital/analog precoders at the transmitter and the combiners at each receiver. By considering receivers with multiple-antenna architecture, our design remarkably improves the overall system performance. Specifically, with only two receive antennas the average transmit power per received message improves by $ 16.8\% $ while the successful information reception is boosted by $ 60\% $. We demonstrate by means of extensive simulations that our hybrid precoder design performs very close to its fully-digital counterpart even under challenging scenarios (i.e., when co-located users belong to distinct multicast groups).      
\end{abstract}

%% file: introduction.tex
\section{Introduction}

In recent years, millimeter-wave (mmWave) has emerged as a promising technology to fuel the ever-increasing consumer demands for extremely fast (i.e., up to multi-Gbps) connectivity. In delivering such requirements for dense networks scenarios (due to extreme densification in next-generation networks), a system can leverage the benefits of multicast communications \cite{b20}. Indeed, recent studies in \cite{b21, b22} demonstrate the potential of multicast to significantly improve the network throughput and spectral efficiency of mmWave systems. To guarantee these performances, an appropriate design for multicast precoders is crucial \emph{(i)} to compensate for severe channel attenuation, and \emph{(ii)} to minimize the interference between simultaneous transmissions. 


An early effort on multicast precoder design is presented in \cite{b1}, where the authors investigated single-group multicast precoding with a multi-antenna base station and several single-antenna receivers. Aiming at minimizing the transmit power subject to predefined quality of service (QoS) requirements (i.e., QoS problem), the problem is formulated as a relaxed semidefinite program (SDP) where befitting solutions are obtained via randomization \cite{b2}. Since their work only considers single-group multicast, the problem formulation thus excludes the interference aspect that is relevant in designing multi-group multicast precoders. Expanding on \cite{b1}, the authors of \cite{b3, b4} investigate a scenario with multiple co-channel multicast groups, which allows transmissions of simultaneous multicast signals by exploiting spatial multiplexing. Furthermore, to mitigate the interference between the distinct signals (and thus increase the number of served users), \cite{b3, b4} an additional stage of power control is incorporated. The QoS problem is also considered in \cite{b5, b6, b7} with diverse extensions to the formulation. A related formulation known as the max-min fair (MMF) problem is studied in \cite{b8, b9, b10}.

The works mentioned above are developed within the framework of fully-digital multicast precoders. Given the affordable hardware and moderate computational complexity of hybrid precoders, a shift of interest has been observed in departing from fully-digital to adopting hybrid antenna arrays architectures. Hybrid precoders are composed of a low-dimensional digital beamformer in cascade with a high-dimensional network of cost-efficient constant-modulus phase shifters that admit a limited number of phase rotations. In general, hybrid precoders are less flexible than their fully-digital counterparts, thus rendering the design of an optimal hybrid precoder a challenging task. Besides, they pose a compromise in terms of beamforming capabilities and interference management. On the other hand, the versatility of digital precoders comes at the expense of highly complicated and expensive hardware, wherein a dedicated radio frequency (RF) chain is required for each antenna element. 

To date, the body of works that has studied hybrid precoding for physical layer multicasting includes single-group multicasting (in \cite{b11}) and multi-group multicasting (in \cite{b12, b13, b14}). The authors of \cite{b11} consider the MMF problem in single-group multicast settings, wherein a codebook-based design is presented. On the other hand, the multi-group multicast QoS and MMF problems are revisited in \cite{b12}, where the authors propose a customized hybrid architecture with improved performance. In \cite{b13}, the authors investigate the QoS problem by considering a high-resolution lens array with adjustable power. However, such design circumvents the constant-modulus discrete phase shifts characteristics of analog circuitry components of hybrid precoders. On the contrary, the authors in \cite{b14} design a scheme to support joint power and information transfer with hybrid precoders. Their formulation considers discrete phase shifts, but restrains the set of phase shifts to only four choices. In addition, the existing studies on multicast precoders for mmWave systems only consider receivers with single-antenna architecture. This conditioning prevents the mitigation of undesired signals (e.g., interference), especially when users from different multicast groups have correlated channel vectors. In particular, endowing receivers with multiple antennas: \emph{(i)} mitigates interference from other sources, \emph{(ii)} reduces the power expenditure from the transmitter, and \emph{(iii)} improves the service ubiquitousness. 

To the best of our knowledge, we are the first to investigate the \emph{joint design of hybrid multicast precoders with an arbitrary number of finite-resolution constant-modulus phase shifts at the transmitter while considering multiple antennas at the receivers}. Related art on hybrid precoding for multi-user scenarios (e.g., in \cite{b15}) are fundamentally different as each RF chain at the transmitter is matched to the channel of one dedicated user. In the multi-group multicast scenario we consider in this paper---due to the limitation of RF chains---several users with distinct channel conditions need to be served by a single RF unit, thus complicating the design of the hybrid precoder. Our proposed formulation focuses on the QoS problem, for which we present an SDR-based approach to optimize the digital precoder, analog phase shifts and receive combiners. Due to the existence of several design parameters, our proposed formulation is divided into a set of sub-problems that we approach adopting alternating optimization (as in e.g., \cite{b17}). Moreover, we incorporate a set of slack parameters to promote coherent parameter binding among the decoupled sub-problems. Since alternating optimization requires each sub-problem to be solvable to guarantee the continuity of the optimization process, such a set of slack parameters ensures that each sub-problem always yields a feasible solution for the succeeding stages. Finally, due to the selection of finite-resolution constant-modulus phase shifts, the problem is inherently of combinatorial nature. To circumvent this matter, we propose a scheme where the phase shifts selection is recast as an SDR program followed by a stage consisting of Cholesky matrix factorization, least squares, and randomization.


The paper is structured as follows. In \sref{sec:system_model}, we model and elaborate on the problem of multi-group multicast hybrid transmit precoders with finite-resolution phase shifts and multi-antenna receivers in mmWave systems. In \sref{sec:problem_formulation}, we formulate the problem and present the proposed solution in \sref{sec:proposed_solution}. We analyze and compare the performance of our design in \sref{sec:result}. Also, we include an insightful discussion in \sref{sec:discussion}. Finally, we conclude with the contributions of this paper in \sref{sec:conclusion}.

%% file: system_model.tex
\section{System Model}
\label{sec:system_model}

We adopt a mmWave system where a gNodeB serves $ K $ users distributed into $ G $ different co-channel multicast groups. The sets of users and groups are denoted  by $ \mathcal{K} =  \left\lbrace 1, 2, \dots, K \right\rbrace $ and $ \mathcal{I} = \left\lbrace 1, 2, \dots, G \right\rbrace $, respectively. Each multicast group $ \mathcal{G}_i \left( i \in \mathcal{I} \right) $ contains the indices of users that constitute it. The amount of users in each multicast group is represented by $ \left| \mathcal{G}_i \right| $, such that $ \sum^G_{i = 1} \left| \mathcal{G}_i \right| = K $. As in \cite{b3}, we assume that $ \mathcal{G}_i \cap \mathcal{G}_j = \left\lbrace  \emptyset \right\rbrace, \forall i \neq j $. The gNodeB is equipped with $ N_\mathrm{tx} $ transmit antennas and $ N^{\mathrm{RF}}_\mathrm{tx} $ RF chains, with $ G \leq N^{\mathrm{RF}}_\mathrm{tx} \leq N_\mathrm{tx} $. The downlink signal is represented by $ \mathbf{x} = \mathbf{F} \mathbf{M} \mathbf{s} $, where $ \mathbf{F} \in \mathbb{C}^{ N_\mathrm{tx} \times N^{\mathrm{RF}}_\mathrm{tx} } $ is the analog precoder whereas $ \mathbf{M} = \left[ \mathbf{m}_1, \mathbf{m}_2, \dots, \mathbf{m}_G \right] \in \mathbb{C}^{N^{\mathrm{RF}}_\mathrm{tx} \times G} $ assembles the digital precoders for each of the multicast group. The collection of data symbols for the intended groups is denoted by $\mathbf{s} = \left[ s_1, s_2, \dots, s_G \right]^T \in \mathbb{C}^{G \times 1}$, where each entry has unit power on average, i.e., $ \mathbb{E} \left\lbrace \mathbf{s} \mathbf{s}^H \right\rbrace = \mathbf{I} $. Also, every element $ \left( q,r \right)  $ of the analog precoder is a phase rotation with constant modulus. Therefore, $ \left[ \mathbf{F} \right]_{q,r} \in \mathcal{F} $, where $ q \in \mathcal{Q} = \left\lbrace 1, 2, \dots, N_\mathrm{tx} \right\rbrace $, $ r \in \mathcal{R} = \left\lbrace 1, 2, \dots, N^{\mathrm{RF}}_\mathrm{tx} \right\rbrace $, $ \mathcal{F} = \left\lbrace \sqrt{\delta}, \sqrt{\delta} e^\frac{2 \pi}{L}, \dots, \sqrt{\delta} e^\frac{2 \pi \left( L-1 \right) }{L} \right\rbrace $, $ L $ denotes the number of different phase shifts that are allowed, and $ \delta $ is a scaling factor. Each multicast receiver has a finite number of receive antennas $ N_\mathrm{rx} \ll N_\mathrm{tx} $, and an equal number of RF chains. Under the assumption of narrowband flat-fading, the signal received by the $ k $-th user ($ k \in \mathcal{G}_i $) is given by
\begin{align} \label{e1}
	\begin{split}
	y_k = \underbrace{
					  			\mathbf{w}^H_k \mathbf{H}_k \mathbf{F} \mathbf{m}_i s_i
		 			 }
		 			 _{\text{desired multicast signal}} +
		  \underbrace{
								\mathbf{w}^H_k \mathbf{H}_k \sum^G_{\substack{j = 1 \\ j \neq i}} \mathbf{F} \mathbf{m}_j s_j
		 			 }
		 			 _{\text{interference}} + 
	 	  \underbrace{
								\mathbf{w}^H_k \mathbf{n}_k
		 			 }
		 			 _{\text{noise}}, 
	\end{split}
\end{align}
where $ i $ is the index of group $ \mathcal{G}_i $, and $ \mathbf{w}_k \in {\mathbb{C}}^{ N_\mathrm{rx} \times 1} $ represents the digital receive beamformer of the $ k $-th user. Also, $ \mathbf{H}_k \in {\mathbb{C}}^{ N_\mathrm{rx} \times N_\mathrm{tx} } $ denotes the channel between the gNodeB and the $ k $-th user, whereas $\mathbf{n}_k \sim \mathcal{CN} \left( \mathbf{0}, {\sigma}^2 \mathbf{I} \right) $ denotes additive white Gaussian noise. The signal--to--interference--plus-noise ratio (SINR) at user $ k $ is defined as
\begin{equation} \label{e2}
	\mathrm{SINR}_k = \frac
			 {\left| \mathbf{w}^H_k \mathbf{H}_k \mathbf{F} \mathbf{m}_i \right|^2}
			 {\sum_{j \neq i} \left| \mathbf{w}^H_k \mathbf{H}_k \mathbf{F} {\mathbf{m}_j} \right|^2 + {\sigma}^2 \left\| \mathbf{w}_k \right\|^2_2}.  
\end{equation}

\section{Problem Formulation}
\label{sec:problem_formulation}

Aiming to optimize the transmit power, we formulate
\begin{subequations} \label{e3}
	\begin{align}
	\mathcal{P}^{\mathrm{hyb}}: & \min_{
								\substack{ 
											\mathbf{F},
											\left\lbrace \mathbf{m}_i \right\rbrace^G_{i=1}, \\
											\left\lbrace \mathbf{w}_k \right\rbrace^K_{k=1}
										 }
							           } 
	& & { \sum^G_{i=1} \left\| \mathbf{F} \mathbf{m}_i \right\|^2_2} \label{e3a} 
	\\
	\vspace{-0.2cm}
	& ~~~~~ \mathrm{s.t.} & &  \frac{\left| \mathbf{w}^H_k \mathbf{H}_k \mathbf{F} \mathbf{m}_i \right|^2}
	{\sum_{j \neq i} \left| \mathbf{w}^H_k \mathbf{H}_k \mathbf{F} {\mathbf{m}_j} \right|^2 + {\sigma}^2 \left\| \mathbf{w}_k \right\|^2_2} \geq \gamma_i, \label{e3b} 
	\\
	& & & \left\| \mathbf{w}_k \right\|^2_2 = P^{\mathrm{max}}_\mathrm{rx}, \label{e3c} 
	\\
	& & & \left\| \mathbf{F} \right\|^2_\mathrm{F} = N^{\mathrm{RF}}_\mathrm{tx}, \label{e3d}
	\\
	& & & \left[ \mathbf{F} \right]_{q,r} \in \mathcal{F}, q \in \mathcal{Q}, r \in \mathcal{R}, \forall k \in \mathcal{G}_i, i\in \mathcal{I}, \label{e3e}
	\end{align}
\end{subequations}
where (\ref{e3a}) targets the minimization of the transmit power. Constraint (\ref{e3b}) imposes specific QoS requirements for each multicast group, whereas (\ref{e3c}) restricts the power expenditure for receive beamforming at each user. Constraint (\ref{e3d}) limits the power associated with the analog precoder. In addition, (\ref{e3e}) enforces every phase shift to belong to $ \mathcal{F} $. The target SINR of every group $ \mathcal{G}_i $ is denoted by $ \gamma_i $. Note that (\ref{e3a}) is non-convex due to multiplicative coupling between $ \mathbf{F} $ and $ \mathbf{m}_i $. Constraint (\ref{e3b}) is non-convex since it is defined as the ratio of two non-convex expressions. On the other hand, (\ref{e3c}) is quadratic and non-convex on $ \mathbf{w}_k $. Constraint (\ref{e3e}) is inherently of combinatorial nature, therefore non-convex. Thus, (\ref{e3d}) is also non-convex due to its dependence on (\ref{e3e}). As a result, $ \mathcal{P}^{\mathrm{hyb}} $ is classified as a non-convex quadratically constrained quadratic program (QCQP), which is known to be NP-hard. 

\section{Proposed Solution}
\label{sec:proposed_solution}

In this section, we propose an approach based on alternating optimization, where the unknown parameters $ \mathbf{F} $, $\left\lbrace \mathbf{m}_i \right\rbrace^G_{i=1}$ and $\left\lbrace \mathbf{w}_k \right\rbrace^K_{k = 1}$ are optimized sequentially and iteratively. Due to sequential (and independent) parameter optimization, the suitability of the solution can be compromised. Therefore, we include an additional set of slack parameters $\left\lbrace x_k \right\rbrace^K_{k = 1}$ to reinforce the linkage between $ \mathbf{F} $, $ \left\lbrace \mathbf{m}_i \right\rbrace^G_{i = 1} $ and $ \left\lbrace \mathbf{w}_k \right\rbrace^K_{k = 1} $. 
Thus, the resulting problem formulation is defined as follows,
\begin{subequations} \label{e4}
	\begin{align}
	\mathcal{P}^{\mathrm{hyb}}_0: & \min_{
									\substack{ 
												\mathbf{F},
												\left\lbrace \mathbf{m}_i \right\rbrace^G_{i=1}, \\
												\left\lbrace \mathbf{w}_k \right\rbrace^K_{k=1}, \\
												\left\lbrace x_k \right\rbrace^K_{k=1}
											 }
								} & & 
	{ \sum^G_{i=1} \left\| \mathbf{F} \mathbf{m}_i \right\|^2_2 + \beta \sum^K_{k = 1} x_k } \label{e4a}
	\\
	\vspace{-0.2cm}
	& ~~~~~ \mathrm{s.t.} & & \frac{\left| \mathbf{w}^H_k \mathbf{H}_k \mathbf{F} \mathbf{m}_i \right|^2 + x_k}
		   {\sum_{j \neq i} \left| \mathbf{w}^H_k \mathbf{H}_k \mathbf{F} {\mathbf{m}_j} \right|^2 + {\sigma}^2 \left\| \mathbf{w}_k \right\|^2_2} \geq \gamma_i, \label{e4b}
	\\
	& & & \left\| \mathbf{w}_k \right\|^2_2 = P^{\mathrm{max}}_\mathrm{rx}, \label{e4c}
	\\
	& & & \left\| \mathbf{F} \right\|^2_\mathrm{F} = N^{\mathrm{RF}}_\mathrm{tx}, \label{e4d}
	\\
	& & & \left[ \mathbf{F} \right]_{q,r} \in \mathcal{F}, \label{e4e}
	\\
	& & & x_k \geq 0, q \in \mathcal{Q}, r \in \mathcal{R}, \forall k \in \mathcal{G}_i, i\in \mathcal{I}. \label{e4f}
	\end{align}
\end{subequations}

Each $ x_k \in \mathbb{R}_{+} $ penalizes the objective function (with a sufficiently large $ \beta $) whenever $ x_k > 0 $ needs to be added to the left-hand side numerator of (\ref{e4b}) for the QoS inequality to hold. Thus, an increment of $ \sum^G_{i=1} \left\| \mathbf{F} \mathbf{m}_i \right\|^2_2 $ will be prioritized instead of letting $ \sum^K_{k = 1} x_k $ augment. This regularization promotes more QoS inequalities to be satisfied by action of $ \mathbf{F} $, $ \left\lbrace \mathbf{m}_i \right\rbrace^G_{i = 1} $ and $ \left\lbrace \mathbf{w}_k \right\rbrace^K_{k = 1} $. The slack parameters $ \left\lbrace x_k \right\rbrace^K_{k=1} $ ensure that a feasible solution always exists as $ x_k $ will absorb any surplus that is required for (\ref{e4b}) to hold. In the following we optimize the three sets of parameters by separating $ \mathcal{P}^{\mathrm{hyb}}_0 $ into 3 sub-problems $ \mathcal{P}^{\mathrm{hyb}}_1 $, $ \mathcal{P}^{\mathrm{hyb}}_2 $ and $ \mathcal{P}^{\mathrm{hyb}}_3 $, which are sequentially and alternately solved. 

\noindent{\textbf{Observation:}} Even with fully-digital precoders and single-antenna receivers (i.e., $ \mathbf{F} = \mathbf{I} $, $ \mathbf{w}_k = 1 $), a feasible solution to (\ref{e4}) cannot always be guaranteed. This usually occurs when $ N_\mathrm{tx} < K $ (as in our case). As a consequence, $ \mathcal{P}^{\mathrm{hyb}}_1 $, $ \mathcal{P}^{\mathrm{hyb}}_2 $ or $ \mathcal{P}^{\mathrm{hyb}}_3 $ may render infeasible, thus interrupting the optimization procedure. To prevent this, we include $ x_k $ to ensure the existence of a feasible solution (without raising infeasibility certificates), thereby guaranteeing the continuity of the sequential optimization process.

\noindent{\textbf{Observation:}} In contrast to adaptive hybrid precoding, where the architectures changes dynamically (i.e., some phase shifters activate/deactivate), in our case the fixed fully-connected architecture allows us to determine $ \delta = 1 / N^\mathrm{RF}_{\mathrm{tx}} $ from (\ref{e4d}). Thus, (\ref{e4d}) is removed in the sequel.

\subsection{Optimization of $\mathbf{F}$}
Assuming that $ \left\lbrace \mathbf{m}_i \right\rbrace^G_{i=1} $ and $ \left\lbrace \mathbf{w}_k \right\rbrace^K_{k=1} $ are known, we optimize over $ \mathbf{F} $. Thus,
\begin{subequations} \label{e5}
	\begin{align}
	{\mathcal{P}}^{\mathrm{hyb}}_1: & \min_{
									\substack{ 
												\mathbf{F},
												\left\lbrace x_k \right\rbrace^K_{k=1}
											 }
								  } & &
	{ \sum^G_{i=1} \left\| \mathbf{F} \mathbf{m}_i \right\|^2_2 + \beta \sum^K_{k=1} x_k } \label{e5a}
	\\
	\vspace{-0.2cm}
	& ~~~~~ \mathrm{s.t.} & & \gamma_i \left( \sum_{j \neq i} \left| \mathbf{w}^H_k \mathbf{H}_k \mathbf{F} {\mathbf{m}_j} \right|^2 + {\sigma}^2 \left\| \mathbf{w}_k \right\|^2_2 \right) \nonumber
	\\
	& & & ~~~~~~~~~~~~~~ - \left| \mathbf{w}^H_k \mathbf{H}_k \mathbf{F} {\mathbf{m}_i} \right|^2  \leq x_k, \label{e5b}
	\\
	& & & \left[ \mathbf{F} \right]_{q,r} \in \mathcal{F}, \label{e5c}
	\\
	& & & x_k \geq 0, q \in \mathcal{Q}, r \in \mathcal{R}, \forall k \in \mathcal{G}_i, i\in \mathcal{I}. \label{e5d}
	\end{align}
\end{subequations}

Notice that we can express $ \mathbf{F} \mathbf{m}_i = \mathbf{J}_i \mathbf{f} $, where $ \mathbf{J}_i = \mathbf{m}^T_i \otimes \mathbf{I} $ and $ \mathbf{f}= \mathrm{vec} \left( \mathbf{F} \right) $. With this redefinition, (\ref{e5}) can be equivalently expressed as,
\begin{subequations} \label{e6}
	\begin{align}
	\mathcal{P}^{\mathrm{hyb}}_1: & \min_{
									\substack{ 
												\mathbf{f},
												\left\lbrace x_k \right\rbrace^K_{k=1}
											 }
								} & & 
	{ \sum^G_{i=1} \left\| \mathbf{J}_i \mathbf{f} \right\|^2_2 + \beta \sum^K_{k=1} x_k } \label{e6a}
	\\
	\vspace{-0.2cm}
	& ~~~~ \mathrm{s.t.} & & \gamma_i \left( \sum_{j \neq i} \left| \mathbf{w}^H_k \mathbf{H}_k \mathbf{J}_j \mathbf{f} \right|^2 + {\sigma}^2 \left\| \mathbf{w}_k \right\|^2_2 \right) \nonumber
	\\
	& & & ~~~~~~~~~~~~~~ - \left| \mathbf{w}^H_k \mathbf{H}_k \mathbf{J}_i \mathbf{f} \right|^2  \leq x_k, \label{e6b}
	\\
	& & & \left[ \mathbf{f} \right]_n \in \mathcal{F}, \label{e6c}
	\\
	& & & x \geq 0, n \in \mathcal{N}, \forall k \in \mathcal{G}_i, i\in \mathcal{I}, \label{e6d}
	\end{align}
\end{subequations}
where $ \mathcal{N} = \left\lbrace 1, 2, \dots, N^{\mathrm{RF}}_\mathrm{tx} N_\mathrm{tx} \right\rbrace $. Note that $ \left\| \mathbf{J}_i \mathbf{f} \right\|^2_2 = \mathrm{Tr} \left( \mathbf{R}_i \mathbf{D} \right) $, with $ \mathbf{R}_i = \mathbf{J}^H_i \mathbf{J}_i $ and $\mathbf{D} = \mathbf{f}\mathbf{f}^H $. Also, $ \left[ \mathbf{D} \right]_{n,n} = \delta $ since $ \left[ \mathbf{f} \right]_n \in \mathcal{F} $. Furthermore, since $ \left| \mathbf{w}^H_k \mathbf{H}_k \mathbf{J}_i \mathbf{f} \right|^2 = \mathrm{Tr} \left( \mathbf{V}_{i,k} \mathbf{D} \right) $, with $ \mathbf{V}_{i,k} = \mathbf{J}^H_i \mathbf{H}^H_k \mathbf{w}_k \mathbf{w}^H_k \mathbf{H}_k \mathbf{J}_i $, we can recast (\ref{e6}) in its SDP form as, 
\begin{subequations} \label{e7}
	\begin{align}
	{\mathcal{P}}^{\mathrm{hyb}}_{\mathrm{SDP},1}: & \min_{
										\substack{ 
											\mathbf{D},
											\left\lbrace x_k \right\rbrace^K_{k=1}
											    }
									  } & &
	{ \sum^G_{i=1} \mathrm{Tr} \left( \mathbf{D} \mathbf{R}_i \right) + \beta \sum^K_{k=1} x_k } \label{e7a}
	\\
	\vspace{-0.2cm}
	& ~~~~~ \mathrm{s.t.} & & \mathrm{Tr} \left( \mathbf{D} \left( \gamma_i \sum_{j \neq i} \mathbf{V}_{j,k} - \mathbf{V}_{i,k} \right) \right) \nonumber
	\\
	& & & ~~~~~~~~~~~ + {\sigma}^2 \gamma_i \left\| \mathbf{w}_k \right\|^2_2 \leq x_k, \label{e7b}
	\\
	& & & \left[ \mathbf{D} \right]_{n,n} = \delta, \label{e7c}
	\\
	& & & \mathrm{rank} \left( \mathbf{D} \right) = 1, \label{e7d}
	\\
	& & & \mathbf{D} \succcurlyeq \mathbf{0}, \label{e7e}
	\\
	& & & x_k \geq 0, n \in \mathcal{N}, \forall k \in \mathcal{G}_i, i\in \mathcal{I}. \label{e7f}
	\end{align}
\end{subequations}

The SDP program in (\ref{e7}) has a linear objective subject to affine constraints except for the non-convex constraint (\ref{e7d}), which imposes a rank-one condition on $ \mathbf{D} $ (as it is originally obtained from $ \mathbf{D} = \mathbf{f} \mathbf{f}^H $). Constraint (\ref{e7e}) restricts $ \mathbf{D} $ to be Hermitian positive semidefinite. It is worth noticing that (\ref{e6c}) is the only constraint not strongly enforced in (\ref{e7}). Thus, while the constant-modulus requirement of (\ref{e6c}) is satisfied by (\ref{e7c}), its phase has been ignored. Nevertheless, the phase will be optimized through the following procedure \cite{b19}. 

\underline{\textit{Stage A\textsubscript{1}:}}
Notice that any element $\left( n_1, n_2 \right) $ of matrix $ \mathbf{D} $ can be represented as $ \left[ \mathbf{D} \right]_{n_1,n_2}  = \left[ \mathbf{f} \right]_{n_1} \left[ \mathbf{f} \right]^{*}_{n_2} $. Now, let us define a vector $ \mathbf{u} \in \mathbb{C}^{ N^{\mathrm{RF}}_\mathrm{tx} N_\mathrm{tx} \times 1} $ such that $ \left\| \mathbf{u}\right\|^2_2 = \mathbf{u}^H \mathbf{u} = 1 $. As a consequence, we can express $ \left[ \mathbf{D} \right]_{n_1,n_2} $ in terms of $ \mathbf{u} $, i.e., $ \left[ \mathbf{D} \right]_{n_1,n_2}  = \left( \left[ \mathbf{f} \right]_{n_1} \mathbf{u}^T \right) \left( \left[ \mathbf{f} \right]^{*}_{n_2} \mathbf{u}^{*} \right) $. Assuming that $ \mathbf{q}_{n} = \left[ \mathbf{f} \right]_{n} \mathbf{u} $, $ \mathbf{D} $ can be recast as $ \mathbf{D} = \mathbf{Q}^T \mathbf{Q}^{*} $ with $ \mathbf{Q} = \left[ \mathbf{q}_1, \mathbf{q}_2, \dots, \mathbf{q}_{ N^{\mathrm{RF}}_\mathrm{tx} N_\mathrm{tx} } \right] $. 

\underline{\textit{Stage A\textsubscript{2}:}}
In (\ref{e7}), the only non-convex constraint is (\ref{e7d}). Thus, we define $ {\mathcal{P}}^{\mathrm{hyb}}_{\mathrm{SDR},1} $ as the resultant SDR surrogate of (\ref{e7}) obtained upon dropping (\ref{e7d}). The solution returned by $ \mathcal{P}^{\mathrm{hyb}}_{\mathrm{SDR},1} $ is denoted by $ \widehat{\mathbf{D}} $. Then, via Cholesky matrix factorization we obtain $ \widehat{\mathbf{D}} = \widehat{\mathbf{Q}}^T \widehat{\mathbf{Q}}^{*} $, where $ \widehat{\mathbf{Q}} = \left[ \widehat{\mathbf{q}}_1, \widehat{\mathbf{q}}_2, \dots, \widehat{\mathbf{q}}_{ N^{\mathrm{RF}}_\mathrm{tx} N_\mathrm{tx} } \right] $. Although we have derived a relation that associates the unknown phase shifts $ \widehat{\mathbf{f}} $ with the known vectors $ \left\lbrace \widehat{\mathbf{q}}_n \right\rbrace^{N^{ \mathrm{RF}}_\mathrm{tx} N_\mathrm{tx} }_{n = 1} $ (via $ \widehat{\mathbf{q}}_{n} = \left[ \widehat{\mathbf{f}} \right]_{n} \widehat{\mathbf{u}} $), the vector $ \widehat{\mathbf{u}} $ also remains unknown. Moreover, the initial premise was that every $ \widehat{\mathbf{q}}_n $ could be obtained from the same $ \widehat{\mathbf{u}} $. However, this cannot be guaranteed as a solution $ \left( \widehat{\mathbf{f}}, \widehat{\mathbf{u}} \right) $ for $ \widehat{\mathbf{q}}_{n} = \left[ \widehat{\mathbf{f}} \right]_{n} \widehat{\mathbf{u}}, \forall n \in \mathcal{N} $ may not exist. Thus, we aim at finding approximate $ \widehat{\mathbf{f}} $ and $ \widehat{\mathbf{u}} $, such that $ \widehat{\mathbf{q}}_{n} \approx \left[ \widehat{\mathbf{f}} \right]_{n} \widehat{\mathbf{u}} $, and whose error is minimum in the 2-norm sense. Mathematically,
\begin{subequations} \label{e8}
	\begin{align}
	\mathcal{P}^{\mathrm{hyb}}_{\mathrm{LS}} : ~~~~~~ & \min_{
									\substack{ 
												\widehat{\mathbf{u}}, 
												\left[ \widehat{\mathbf{f}} \right]_{n}
											 }
								   } & &
	{ 
		\sum^{ N^{\mathrm{RF}}_\mathrm{tx} N_\mathrm{tx} }_{n = 1} \left\| \widehat{\mathbf{q}}_n - \left[ \widehat{\mathbf{f}} \right]_n  \widehat{\mathbf{u}} \right\|^2_2
	} \label{e8a}
	\\
	\vspace{-0.2cm}
	& ~~ \mathrm{s.t.} & & \left\| \widehat{\mathbf{u}} \right\|^2_2 = 1, ~~~~~~~~~~~~~~~~~~~~~~~~~~~~ \label{e8b}
	\\
	& & & \left[ \widehat{\mathbf{f}} \right]_n \in \mathcal{F}, n \in \mathcal{N}. \label{e8c}
	\end{align}
\end{subequations}

\underline{\textit{Stage A\textsubscript{3}:}} 
Minimizing simultaneously over both $ \widehat{\mathbf{q}}_n $ and $ \widehat{\mathbf{u}} $ is challenging. If we assume that $ \widehat{\mathbf{u}} $ is known such that (\ref{e8b}) is satisfied, then we are required to solve 
\begin{subequations} \label{e9}
	\begin{align}
	\widetilde{\mathcal{P}}^{\mathrm{hyb}}_{\mathrm{LS}} : ~~~~~~ & \min_{
												\substack{ 
															\left[ \widehat{\mathbf{f}} \right]_{n}
														 }
											   } & &
	{ 
		\sum^{ N^{\mathrm{RF}}_\mathrm{tx} N_\mathrm{tx} }_{n = 1} \left\| \widehat{\mathbf{q}}_n - \left[ \widehat{\mathbf{f}} \right]_n  \widehat{\mathbf{u}} \right\|^2_2
	} \label{e9a}
	\\
	\vspace{-0.2cm}
	& ~ \mathrm{s.t.} & & \left[ \widehat{\mathbf{f}} \right]_n \in \mathcal{F}, n \in \mathcal{N} ~~~~~~~~~~~~~~~~~~~~~~ \label{e9b}
	\end{align}
\end{subequations}

By expanding (\ref{e9a}), we realize that $ \left\| \widehat{\mathbf{q}}_n - \left[ \widehat{\mathbf{f}} \right]_n  \widehat{\mathbf{u}} \right\|^2_2 = \widehat{\mathbf{q}}^H_n \widehat{\mathbf{q}}_n - 2 \mathfrak{Re} \left( \left[ \widehat{\mathbf{f}} \right]_n \widehat{\mathbf{q}}^H_n \widehat{\mathbf{u}} \right) + \left| \left[ \widehat{\mathbf{f}} \right]_n \right|^2 \widehat{\mathbf{u}}^H \widehat{\mathbf{u}} $. Thus, (\ref{e9}) is equivalent to
\begin{subequations} \label{e10}
	\begin{align}
	\widetilde{\mathcal{P}}^{\mathrm{hyb}}_{\mathrm{LS}} : ~~~~~~ & \max_{
												\substack{ 
															\left[ \widehat{\mathbf{f}} \right]_{n}
														 }
											   } & &
	{ 
		\sum^{ N^{\mathrm{RF}}_\mathrm{tx} N_\mathrm{tx} }_{n = 1} \mathfrak{Re} \left( \left[ \widehat{\mathbf{f}} \right]_n \widehat{\mathbf{q}}^H_n \widehat{\mathbf{u}} \right)
	} \label{e10a}
	\\
	\vspace{-0.2cm}
	& ~~ \mathrm{s.t.} & & \left[ \widehat{\mathbf{f}} \right]_n \in \mathcal{F}, n \in \mathcal{N}. ~~~~~~~~~~~~~~~~~~~~~~ \label{e10b}
	\end{align}
\end{subequations}

Note that (\ref{e10}) can be decomposed into $ N^{\mathrm{RF}}_\mathrm{tx} N_\mathrm{tx} $ independent sub-problems. Thus, since $ z_n = \widehat{\mathbf{q}}^H_n \widehat{\mathbf{u}} $ is known, we need to select $ \left[ \mathbf{f} \right]_n $ such that the real part of (\ref{e10a}) is maximized. This is equivalent to choosing $ \left[ \widehat{\mathbf{f}} \right]_n \in \mathcal{F} $ with the closest phase to $ z^{*}_n $. After finding $ \widehat{\mathbf{f}} $, it can be reshaped in order to obtain $ \widehat{\mathbf{F}} $. As shown in Algorithm \ref{a1}, $ N_\mathrm{rand} $ candidate vectors $ \widehat{\mathbf{u}} $ are generated and the best-performing option is maintained.

\subsection{Optimization of $ \mathbf{m}_i $}
We assume herein that $ \mathbf{F} $ and $ \left\lbrace \mathbf{w}_k \right\rbrace^K_{k=1} $ are known. Thus, the original problem in (\ref{e4}) collapses to
\begin{subequations} \label{e11}
	\begin{align}
	\mathcal{P}^{\mathrm{hyb}}_2: & \min_{
									\substack{ 
												\left\lbrace \mathbf{m}_i \right\rbrace^G_{i=1}, \\
												\left\lbrace x_k \right\rbrace^K_{k=1}
											 }
							    } & &
	{ \sum^G_{i=1} \left\| \mathbf{F} \mathbf{m}_i \right\|^2_2 + \beta \sum^K_{k=1} x_k } \label{e11a}
	\\
	\vspace{-0.2cm}
	& ~~~~ \mathrm{s.t.} & & \gamma_i \left( \sum_{j \neq i} \left| \mathbf{w}^H_k \mathbf{H}_k \mathbf{F} {\mathbf{m}_j} \right|^2 + {\sigma}^2 \left\| \mathbf{w}_k \right\|^2_2 \right) \nonumber
	\\
	& & & ~~~~~~~~~~~~~~ - \left| \mathbf{w}^H_k \mathbf{H}_k \mathbf{F} {\mathbf{m}_i} \right|^2  \leq x_k, \label{e11b}
	\\
	& & & x_k \geq 0, \forall k \in \mathcal{G}_i, i\in \mathcal{I}. \label{e11c}
	\end{align}
\end{subequations}

The SDP equivalent formulation of (\ref{e11}) is expressed as
\begin{subequations} \label{e12}
	\begin{align}
	\mathcal{P}^{\mathrm{hyb}}_{\mathrm{SDP},2}: & \min_{
										\substack{ 
													\left\lbrace \mathbf{M}_i \right\rbrace^G_{i=1}, \\
													\left\lbrace x_k \right\rbrace^K_{k=1}
											 	}
									  } & &
	{ 
		\sum^G_{i = 1}  \mathrm{Tr} \left( \mathbf{Y} \mathbf{M}_i \right) +
		\beta \sum^K_{k=1} x_k 
	} \label{e12a}
	\\
	\vspace{-0.2cm}
	& ~~~~ \mathrm{s.t.} & & \mathrm{Tr} \left( \mathbf{X}_k \left( \gamma_i \sum_{j \neq i} \mathbf{M}_j - \mathbf{M}_i \right) \right) \nonumber
	\\
	& & & ~~~~~~~~~~ + {\sigma}^2 \gamma_i \left\| \mathbf{w}_k \right\|^2_2 \leq x_k, \label{e12b}
	\\
	& & & \mathbf{M}_i \succcurlyeq \mathbf{0}, \label{e12c}
	\\
	& & & \mathrm{rank} \left( \mathbf{M}_i \right) = 1, \label{e12d}
	\\
	& & & x_k \geq 0, \forall k \in \mathcal{G}_i, i\in \mathcal{I}, \label{e12e}
	\end{align}
\end{subequations}
where $ \mathbf{Y} = \mathbf{F}^H \mathbf{F} $,  $ \mathbf{X}_k = \mathbf{F}^H \mathbf{H}^H_k \mathbf{w}_k \mathbf{w}^H_k \mathbf{H}_k \mathbf{F} $ and $ \mathbf{M}_i = \mathbf{m}_i \mathbf{m}^H_i $. Similarly as before, (\ref{e12}) has a linear objective with affine constraints except for (\ref{e12d}). Thus, we define $ \mathcal{P}^{\mathrm{hyb}}_{\mathrm{SDR},2} $ as the SDR surrogate of (\ref{e12}), where (\ref{e12d}) is neglected.

\subsection{Optimization of $\mathbf{w}_k$}
Now, we assume that $ \mathbf{F} $ and $ \left\lbrace \mathbf{m}_i \right\rbrace^G_{i=1} $ are given. Therefore, we optimize over $ \left\lbrace \mathbf{w}_k \right\rbrace^K_{k=1} $ as shown in (\ref{e13})
\begin{subequations} \label{e13}
	\begin{align}
	\mathcal{P}^{\mathrm{hyb}}_3: & \min_{
									\substack{ 
												\left\lbrace \mathbf{w}_k \right\rbrace^K_{k = 1}, \\
												\left\lbrace x_k \right\rbrace^K_{k = 1}
											 }
								} & &
	{ \sum^K_{k = 1} x_k } \label{e13a}
	\\
	\vspace{-0.2cm}
	& ~~~~ \mathrm{s.t.} & & \gamma_i \left( \sum_{j \neq i} \left| \mathbf{w}^H_k \mathbf{H}_k \mathbf{F} {\mathbf{m}_j} \right|^2 + {\sigma}^2 \left\| \mathbf{w}_k \right\|^2_2 \right) \nonumber
	\\
	& & & ~~~~~~~~~~~~~~ - \left| \mathbf{w}^H_k \mathbf{H}_k \mathbf{F} \mathbf{m}_i \right|^2 \leq x_k, \label{e13b}
	\\
	& & & \left\| \mathbf{w}_k \right\|^2_2 = P^{\mathrm{max}}_\mathrm{rx}, \label{e13c}
	\\
	& & & x_k \geq 0, \forall k \in \mathcal{G}_i, i\in \mathcal{I}, \label{e13d}
	\end{align}
\end{subequations}

In SDP form, (\ref{e13}) can be recast as
\begin{subequations} \label{e14}
	\begin{align}
	\mathcal{P}^{\mathrm{hyb}}_{\mathrm{SDP},3}: & \min_{
										\substack{ 
													\left\lbrace \mathbf{W}_k \right\rbrace^K_{k = 1}, \\
													\left\lbrace x_k \right\rbrace^K_{k = 1}
												 }
									  } & &
	{ \sum^K_{k = 1} x_k } \label{e14a}
	\\
	\vspace{-0.2cm}
	& ~~~~ \mathrm{s.t.} & & \mathrm{Tr} \left( \mathbf{W}_k \left( \gamma_i \sum_{j \neq i} \mathbf{Z}_{k,j} - \mathbf{Z}_{k,i} \right) \right) \nonumber
	\\
	& & & ~~~~~~~~~~~~ + {\sigma}^2 \gamma_i \mathrm{Tr} \left( \mathbf{W}_k \right) \leq x_k, \label{e14b}
	\\
	& & & \mathrm{Tr} \left( \mathbf{W}_k \right) = P^{\mathrm{max}}_\mathrm{rx},\label{e14c}
	\\
	& & & \mathbf{W}_k \succcurlyeq \mathbf{0}, \label{e14d}
	\\
	& & & \mathrm{rank} \left( \mathbf{W}_k \right) = 1, \label{e14e}
	\\
	& & & x_k \geq 0, \forall k \in \mathcal{G}_i, i\in \mathcal{I}, \label{e14f}
	\end{align}
\end{subequations}
where $ \mathbf{W}_k = \mathbf{w}_k \mathbf{w}^H_k $ and $ \mathbf{Z}_{k,i} = \mathbf{H}_k \mathbf{F} \mathbf{m}_i \mathbf{m}^H_i \mathbf{F}^H \mathbf{H}^H_k $. Now, we define $ \mathcal{P}^{\mathrm{hyb}}_{\mathrm{SDR},3} $ as (\ref{e14}) without the non-convex constraint (\ref{e14e}). Furthermore, since the optimization of $ \mathbf{W}_k $ only affects $ \mathrm{SINR}_k $, then $ \mathcal{P}^{\mathrm{hyb}}_{\mathrm{SDR},3} $ can be split into $ K $ parallel sub-problems $ \mathcal{P}^{\mathrm{hyb}}_{\mathrm{SDR},3, k} $. For completeness, we summarize our proposed scheme in Algorithm \ref{a1} with more implementation details. Note that $ {\mathcal{P}}^{\mathrm{hyb}}_{\mathrm{SDR},1} $, $ \mathcal{P}^{\mathrm{hyb}}_{\mathrm{SDR},2} $ and $ \mathcal{P}^{\mathrm{hyb}}_{\mathrm{SDR},3} $ can be recast as linear programs and can therefore be efficiently solved in polynomial time by numerical solvers. In our case, we employed \texttt{CVX} and \texttt{SDPT3}. In Algorithm \ref{a1}, $ g^{(t)} $ computes the total transmit power, whereas $ \mathcal{K}^{(t)} $ counts the number of users whose QoS requirement has been satisfied at iteration $ t $. In the initialization stage, every $ \mathbf{w}_k $ is set in omnidirectional reception mode, i.e., only one antenna is active. 
Similarly, every multicast precoder $ \mathbf{m}_i $ is in omnidirectional mode. Then, $ \mathbf{F} $, $ \left\lbrace \mathbf{m}_i \right\rbrace^{G}_{i = 1} $ and $ \left\lbrace \mathbf{w}_k \right\rbrace^K_{k = 1} $ are alternately optimized for a number of iterations $ N_{\mathrm{iter}} $. At each iteration, the SDR-based solutions are used to generate $ N_{\mathrm{rand}} $ potentially befitting solutions.
\begin{algorithm} [!t]
\caption{Proposed Iterative Approach }
	\label{a1}
	\footnotesize
	\textbf{Define} \\
	\begin{tabular}{m{7.9cm}}
		Let $ g^{(t)} = \sum^G_{i=1} \left\| \mathbf{F}^{(t)} \mathbf{m}^{(t)}_i \right\|^2_2 $ be the total transmit power. \\
		Let $ \mathcal{K}^{(t)}$ be the number of users that satisfy (\ref{e5b}) at iteration $ t $. \\
	\end{tabular}
	\textbf{Initialize} \\
	\begin{tabular}{m{7.9cm}}
		Set $\mathbf{w}^{(0)}_k \leftarrow \left[ 1 ~ \mathbf{0} \right]^T, \forall k \in \mathcal{K} $, $ \mathbf{m}^{(0)}_i \leftarrow \left[ 1 ~ \mathbf{0} \right]^T, \forall i \in \mathcal{I} $. \\
		Set $ \widetilde{\mathcal{K}} \leftarrow 0$, $\tilde{g} \leftarrow 10^5$, $t \leftarrow 1 $.
	\end{tabular}
	
	\textbf{Iterate}\\
	{
		\begin{tabular}{m{8.5cm}}
			Set $ C_1 \leftarrow 0 $, $ C_2 \leftarrow 0 $ and $ \left\lbrace C_{3,k} \right\rbrace^K_{k = 1} \leftarrow 0 $. \\
			\underline{Optimize $ \mathbf{F} $:} \\
			~~~~ Solve $ {\mathcal{P}}^{\mathrm{hyb}}_{\mathrm{SDR},1} $ to obtain $ \mathbf{D}^{(t)} $. \\
			~~~~ \textbf{repeat} \\
			~~~~~~~~ Generate $ \mathbf{u} $ with uniform distribution in the sphere $ \left\| \mathbf{u} \right\|^2_2  = 1 $. \\
			~~~~~~~~ Solve $ \widetilde{\mathcal{P}}^{\mathrm{hyb}}_{\mathrm{LS}} $ and compute $ \mathbf{F}^{(t)} $. \\
			~~~~~~~~ \textbf{if} $ \mathcal{K}^{(t)} > \widetilde{\mathcal{K}} $ or $ \left( \mathcal{K}^{(t)} = \widetilde{\mathcal{K}} \text{ and } g^{(t)} \leq \tilde{g} \right) $ \\
			~~~~~~~~~~~~ Assign $ \mathbf{F} \leftarrow \mathbf{F}^{(t)} $, $ \tilde{g} \leftarrow g^{(t)} $, $ \widetilde{\mathcal{K}} \leftarrow \mathcal{K}^{(t)} $. \\
			~~~~~~~~ \textbf{end} \\
			~~~~~~~~ Increase the counter $ C_1 $, $ C_1 \leftarrow C_1 + 1 $. \\
			~~~~ \textbf{while} $ C_1 \leq N_{\mathrm{rand}} $ \\
			
			\underline{Optimize $ \mathbf{m}_i $:} \\
			~~~~ Solve $ \mathcal{P}^{\mathrm{hyb}}_{\mathrm{SDR},2} $ and obtain $ \left\lbrace \mathbf{M}^{(t)}_i \right\rbrace_{i = 1}^G $. \\
			~~~~ \textbf{repeat} \\
			~~~~~~~~ Generate $ \widetilde{\mathbf{m}}^{(t)}_i \sim \mathcal{CN} \left( \mathbf{0}, \mathbf{M}^{(t)}_i \right), \forall i \in \mathcal{I} $. \\
			~~~~~~~~ \textbf{if} $ \mathcal{K}^{(t)} > \widetilde{\mathcal{K}} $ or $ \left( \mathcal{K}^{(t)} = \widetilde{\mathcal{K}} \text{ and } g^{(t)} \leq \tilde{g} \right) $ \\
			~~~~~~~~~~~~ Assign $ \left\lbrace \mathbf{m}_i \right\rbrace_{i=1}^G \leftarrow \left\lbrace \mathbf{m}^{(t)}_i \right\rbrace_{i = 1}^G $, $ \tilde{g} \leftarrow g^{(t)} $, $ \widetilde{\mathcal{K}} \leftarrow \mathcal{K}^{(t)} $. \\
			~~~~~~~~ \textbf{end} \\
			~~~~~~~~ Increase the counter $ C_2 $, $ C_2 \leftarrow C_2 + 1 $. \\
			~~~~ \textbf{while} $ C_2 \leq N_{\mathrm{rand}} $ \\	
			
			\underline{Optimize $ \mathbf{w}_k $:} \\
			~~~~ Solve $ \mathcal{P}^{\mathrm{hyb}}_{\mathrm{SDR},3} $ and obtain $ \left\lbrace \mathbf{W}^{(t)}_k \right\rbrace_{k = 1}^K $. \\
			~~~~ \textbf{repeat for each} $ k $ \\
			~~~~~~~~ Generate $ \mathbf{w}^{(t)}_k \leftarrow \mathbf{W}^{(t)}_k \mathbf{v}_k, \forall k \in \mathcal{K} $ with $ \mathbf{v}_k $ uniformly \\
			~~~~~~~~ distributed in the sphere $ \left\| \mathbf{v}_k \right\|^2_2  = 1 $. \\
			~~~~~~~~ \textbf{if} $ \mathcal{K}^{(t)} > \widetilde{\mathcal{K}} $ or $ \left( \mathcal{K}^{(t)} = \widetilde{\mathcal{K}} \text{ and } g^{(t)} \leq \tilde{g} \right) $ \\
			~~~~~~~~~~~~ Assign $ \mathbf{w}_k \leftarrow \mathbf{w}^{(t)}_k $, $ \tilde{g} \leftarrow g^{(t)} $, $ \widetilde{\mathcal{K}} \leftarrow \mathcal{K}^{(t)} $ \\
			~~~~~~~~ \textbf{end} \\
			~~~~~~~~~~~~~~ Increase the counter $ C_{3,k} $, $ C_{3,k} \leftarrow C_{3,k} + 1 $ \\
			~~~~ \textbf{while} $ C_{3,k} \leq \lfloor N_{\mathrm{rand}} / K \rfloor $ \\	
		\end{tabular}	
	} \\
	\textbf{Until} $ t > N_{\mathrm{iter}} $
\end{algorithm}

%% file: result.tex
\section{Simulation Results}
\label{sec:result}

To evaluate our proposed design, we consider the geometric channel model with $ M_p = 8 $ propagation paths between the transmitter and each user. The maximum receive power for all the users is $ P^{\mathrm{max}}_\mathrm{rx} = 10 $ dBm, and $ \mathcal{F} $ consists of $ L = 8 $ different phase shifts equally spaced in a circle with radius $ \sqrt \delta $. The regularization hyper-parameter in (\ref{e4}) is defined as $ \beta = G^3 N^{\mathrm{RF}}_\mathrm{tx} N_\mathrm{tx} N_\mathrm{rx} $. In the following scenarios, we compare the performance of fully-digital and hybrid precoders in terms of the number of decoded packets $ N_\mathrm{packets} $ and the required transmit power $ P_\mathrm{tx} $ for several configurations of $ \gamma_i $, $ N^{\mathrm{RF}}_\mathrm{tx} $, $ N_\mathrm{tx} $, $ N_\mathrm{rx} $, $ N_{\mathrm{rand}} $ and $ N_{\mathrm{iter}} $. In the sequel, we consider $ K = 60 $ users evenly distributed among $ G = 4 $ multicast groups. All groups have the same SINR requirements, i.e., $ \gamma = \gamma_i \left(\forall i \in \mathcal{I} \right) $ and $ \sigma^2 = 10 $ dBm. All the numerical results show the average over 100 channel realizations.

\subsubsection{Impact of $ N^{\mathrm{RF}}_\mathrm{tx} $} The objective of this experiment is to evaluate the performance of the hybrid precoder with respect to its fully-digital counterpart, when $ N^{\mathrm{RF}}_\mathrm{tx} = \left\lbrace 5, 6, 7, 8, 9, 10, 11 \right\rbrace $ is varied for different $ \gamma = \left\lbrace 4, 6, 8 \right\rbrace $. We assume that $ N_\mathrm{tx} = 12 $, $ N_\mathrm{rx} = 2 $, $ N_\mathrm{iter} = 3 $ and $ N_{\mathrm{rand}} = 500 $. The results for this setting are shown in Fig. \ref{f1}, where the hybrid precoder is denoted by \texttt{HY} and the fully-digital by \texttt{FD}. We observe that for any specific $ \gamma $, the number of decoded packets $ N_\mathrm{packets} $ augments when $ N_{\mathrm{RF}} $ increases. By observing $ N_\mathrm{packets} $, it is evident that it suffices to only have $ N^{\mathrm{RF}}_\mathrm{tx} = 8 $ to yield a similar performance as that of the fully-digital precoder (which requires $ N^{\mathrm{RF}}_\mathrm{tx} = 12 $). However, due to the reduced number of RF chains in the hybrid precoder, interference management becomes more challenging at the transmitter. Thus, we observe that in general, the hybrid precoder requires more transmit power $ P_\mathrm{tx} $ to attain a similar performance. As more RF chains are added, the required transmit power decreases as interference can be more effectively mitigated. It is also worth noting an apparently abnormal behavior that, for instance, occurs when $ \gamma = 6 $ dB for $ N^{\mathrm{RF}}_\mathrm{tx} = 7 $ and $ N^{\mathrm{RF}}_\mathrm{tx} = 8 $. Observe that $ P_\mathrm{tx} $ is higher when $ N^{\mathrm{RF}}_\mathrm{tx} = 8 $ than when $ N^{\mathrm{RF}}_\mathrm{tx} = 7 $. However, $ N_\mathrm{packets} $ is larger by (approximately) one unit when $ N^{\mathrm{RF}}_\mathrm{tx} = 8 $. The reason is that some users experiencing high interference cannot be served when $ N^{\mathrm{RF}}_\mathrm{tx} = 7 $. Nevertheless, when an additional RF chain is incorporated, the number of degrees of freedom increases and oftentimes a subset of the uncatered users can be served at the expense of boosting the transmit power. The maximum value of $ N_\mathrm{packets} $ is $ 60 $ as we consider one transmitted message per user. Due to the highly interfering scenario we have considered, not even the fully-digital precoder with $ N_{\mathrm{rand}} = 500 $ can guarantee $ 100 $\% successful reception.
\begin{figure}[t]
	\hspace{-3mm}
	\centering
	\begin{tikzpicture}[inner sep = 0.3mm]
	\begin{axis}[
	ybar,
	ymin = 30,
	ymax = 60,
	width = 9.25cm,
	height = 3.7cm,
	bar width = 6pt,
	tick align = inside,
	ytick style = {draw = none},
	x label style={align=center, font=\footnotesize,},
	ylabel = {$ N_\mathrm{packets} $},
	y label style={at={(-0.075,0.5)}, font=\footnotesize,},
	nodes near coords,
	every node near coord/.append style={yshift = -1pt, color = black, rotate = 90, anchor = east, font = \fontsize{2}{2}\selectfont},
	nodes near coords align = {vertical},
	symbolic x coords = {$ \gamma = 4~dB $, $ \gamma = 6~dB $, $ \gamma = 8~dB $},
	x tick label style = {text width = 2cm, align = center, font = \fontsize{7}{8}\selectfont},
	y tick label style = {font = \fontsize{7}{8}\selectfont},
	xtick = data,
	enlarge y limits = {value = 0.0, upper},
	enlarge x limits = 0.24,
	legend columns = 8,
	legend pos = north east,
	]
	
	\addplot[fill = gcolor1] coordinates {($ \gamma = 4~dB $, 58.4900) ($ \gamma = 6~dB $, 58.3400) ($ \gamma = 8~dB $, 58.3800)}; 
	
	\addplot[draw=black, fill = white, postaction={pattern = horizontal lines, pattern color = bcolor8}, every node near coord/.append style={fill = white, opacity = 1}] coordinates {($ \gamma = 4~dB $, 49.3500) ($ \gamma = 6~dB $, 48.6600) ($ \gamma = 8~dB $, 41.9000)}; 
	
	\addplot[draw=black, fill = white, postaction={pattern = dots, pattern color = bcolor8}, every node near coord/.append style={fill = white, opacity = 1}] coordinates {($ \gamma = 4~dB $, 53.7000) ($ \gamma = 6~dB $, 51.3000)($ \gamma = 8~dB $, 45.7600)}; 
	
	\addplot[draw=black, fill = white, postaction={pattern = north east lines, pattern color = bcolor8}, every node near coord/.append style={fill = white, opacity = 1}] coordinates {($ \gamma = 4~dB $, 56.7200) ($ \gamma = 6~dB $, 55.5600) ($ \gamma = 8~dB $, 51.9400)}; 
	
	\addplot[draw=black, fill = white, postaction={pattern = north west lines, pattern color = bcolor8}, every node near coord/.append style={fill = white, opacity = 1}] coordinates {($ \gamma = 4~dB $, 57.1400) ($ \gamma = 6~dB $, 56.4900) ($ \gamma = 8~dB $, 53.2900)}; 
	
	\addplot[draw=black, fill = white, postaction={pattern = grid, pattern color = bcolor8}, every node near coord/.append style={fill = white, opacity = 1}] coordinates {($ \gamma = 4~dB $, 57.6700) ($ \gamma = 6~dB $, 57.0400) ($ \gamma = 8~dB $, 56.6700)}; 
	
	\addplot[draw=black, fill = white, postaction={pattern = crosshatch, pattern color = bcolor8}, every node near coord/.append style={fill = white, opacity = 1}] coordinates {($ \gamma = 4~dB $, 57.8900) ($ \gamma = 6~dB $, 57.6900) ($ \gamma = 8~dB $, 56.9400)}; 
	
	\addplot[draw=black, fill = white, postaction={pattern = crosshatch dots, pattern color = bcolor8}, every node near coord/.append style={fill = white, opacity = 1}] coordinates {($ \gamma = 4~dB $, 58.2100) ($ \gamma = 6~dB $, 58.1500) ($ \gamma = 8~dB $, 57.9600)}; 
	
	\end{axis}
	\end{tikzpicture}
	
	\centering
	\begin{tikzpicture}[inner sep = 0.3mm]
	\begin{axis}[
	ybar,
	ymin = 10,
	ymax = 40,
	width = 9.25cm,
	height = 3.7cm,
	bar width = 6pt,
	tick align = inside,
	xtick style = {draw = none},
	x label style={align=center, font=\footnotesize,},
	ylabel = {$ P_\mathrm{tx} $ [dBm]},
	y label style={at={(-0.075,0.5)}, font=\footnotesize,},
	nodes near coords,
	every node near coord/.append style={yshift = -1pt, color = black, rotate = 90, anchor = east, font = \fontsize{2}{2}\selectfont},
	nodes near coords align = {vertical},
	symbolic x coords = {$ \gamma = 4~dB $,  $ \gamma = 6~dB $, $ \gamma = 8~dB $},
	x tick label style = {text width = 2cm, align = center, font = \fontsize{7}{8}\selectfont},
	y tick label style = {font = \fontsize{7}{8}\selectfont},
	xtick = data,
	enlarge y limits = {value = 0.0, upper},
	enlarge x limits = 0.24,
	legend columns = 4,
	legend pos = north east,
	legend style={at={(-0.015,-0.6)},anchor=south west, font=\fontsize{6}{5}\selectfont, text width=1.56cm,text height=0.02cm,text depth=.ex, fill = none}
	]
	
	\addplot[fill = gcolor1] coordinates {($ \gamma = 4~dB $, 24.1131) ($ \gamma = 6~dB $, 27.6724) ($ \gamma = 8~dB $, 32.2303)}; \addlegendentry{ \texttt{FD} $ | N_{\mathrm{RF}} = 12 $ }
	
	\addplot[draw=black, fill = white, postaction={pattern = horizontal lines, pattern color = bcolor8}, every node near coord/.append style={fill = white, opacity = 1}] coordinates {($ \gamma = 4~dB $, 31.1028) ($ \gamma = 6~dB $, 32.7909) ($ \gamma = 8~dB $, 31.4436)}; \addlegendentry{ \texttt{HY} $ | N_{\mathrm{RF}} = 5 $}
	
	\addplot[draw=black, fill = white, postaction={pattern = dots, pattern color = bcolor8}, every node near coord/.append style={fill = white, opacity = 1}] coordinates {($ \gamma = 4~dB $, 34.5183) ($ \gamma = 6~dB $, 37.0656) ($ \gamma = 8~dB $, 35.6890)}; \addlegendentry{ \texttt{HY} $ | N_{\mathrm{RF}} = 6 $}
	
	\addplot[draw=black, fill = white, postaction={pattern = north east lines, pattern color = bcolor8}, every node near coord/.append style={fill = white, opacity = 1}] coordinates {($ \gamma = 4~dB $, 32.3494) ($ \gamma = 6~dB $, 33.8064) ($ \gamma = 8~dB $, 35.9428)}; \addlegendentry{ \texttt{HY} $ | N_{\mathrm{RF}} = 7 $}
	
	\addplot[draw=black, fill = white, postaction={pattern = north west lines, pattern color = bcolor8}, every node near coord/.append style={fill = white, opacity = 1}] coordinates {($ \gamma = 4~dB $, 30.6989) ($ \gamma = 6~dB $, 37.4251) ($ \gamma = 8~dB $, 38.5530)}; \addlegendentry{ \texttt{HY} $ | N_{\mathrm{RF}} = 8 $}	 
	
	\addplot[draw=black, fill = white, postaction={pattern = grid, pattern color = bcolor8}, every node near coord/.append style={fill = white, opacity = 1}] coordinates {($ \gamma = 4~dB $, 29.8004) ($ \gamma = 6~dB $, 31.5818) ($ \gamma = 8~dB $, 36.1673)}; \addlegendentry{ \texttt{HY} $ | N_{\mathrm{RF}} = 9 $}	 
	
	\addplot[draw=black, fill = white, postaction={pattern = crosshatch, pattern color = bcolor8}, every node near coord/.append style={fill = white, opacity = 1}] coordinates {($ \gamma = 4~dB $, 28.0450) ($ \gamma = 6~dB $, 32.1555) ($ \gamma = 8~dB $, 34.2214)}; \addlegendentry{ \texttt{HY} $ | N_{\mathrm{RF}} = 10 $}	 
	
	\addplot[draw=black, fill = white, postaction={pattern = crosshatch dots, pattern color = bcolor8}, every node near coord/.append style={fill = white, opacity = 1}] coordinates {($ \gamma = 4~dB $, 25.7002) ($ \gamma = 6~dB $, 30.5933) ($ \gamma = 8~dB $, 34.7112)}; \addlegendentry{ \texttt{HY} $ | N_{\mathrm{RF}} = 11 $}	   
	
	\end{axis}
	\end{tikzpicture}
	\vspace{-1mm}
	\caption{Evaluation of the number of decoded packets and transmit power for $ N_\mathrm{tx} = 12 $ when $ \gamma $ and $ N^{\mathrm{RF}}_\mathrm{tx} $ are varied.}
	\label{f1}
	\vspace{-3mm}
\end{figure}
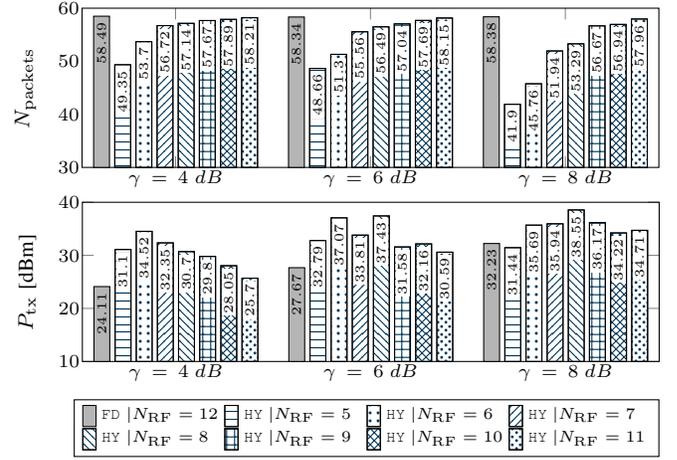

\begin{figure}[t]
	\centering
	\begin{tikzpicture}[inner sep = 0.3mm]
	\begin{groupplot}[
	group style = {group size=2 by 1},
	ybar,
	width = 5cm,
	height = 3.7cm,
	/tikz/bar width = 7pt,
	tick align = inside,
	nodes near coords,
	every node near coord/.append style={yshift = -2pt, color = black, rotate = 90, anchor = east, font = \fontsize{1}{2}\selectfont},
	nodes near coords align = {vertical},
	symbolic x coords = {$ 1 $, $ 2 $, $ 3 $, $ 4 $, $ 5 $},
	x tick label style = {text width = 2cm, align = center, font = \fontsize{7}{8}\selectfont},
	y tick label style = {font = \fontsize{7}{8}\selectfont},
	xtick = data,
	enlarge y limits = {value = 0.0, upper},
	enlarge x limits = 0.12,
	legend columns = 8,
	legend pos = north east,
	legend style={at={(0, 1.01)}, anchor=south west, font=\fontsize{6}{5}\selectfont, text width=1.55cm,text height=0.02cm,text depth=.ex, fill = none, draw = none}
	]
	
\nextgroupplot[
	ymin = 20,
	ymax = 62,
	ylabel = {$ N_\mathrm{packets} $},
	y label style={at={(-0.12,0.5)}, font=\footnotesize,},
	xlabel = {$ N_\mathrm{rx} $},
	x label style={at={(0.5,-0.1)}, font=\footnotesize,}
	]
	
	\addlegendimage{fill = gcolor1, mark=none, line width=0.5pt},
	\addlegendentry{Fully-digital (\texttt{FD})},
	\addlegendimage{fill = bcolor8, mark=none, line width=0.5pt, color = black, pattern = crosshatch, pattern color = bcolor8},
	\addlegendentry{Hybrid (\texttt{HY})},
	
	\addplot[fill = gcolor1] coordinates {($ 1 $, 54.2400) ($ 2 $, 59.1600) ($ 3 $, 59.7000) ($ 4 $, 59.9200) ($ 5 $, 59.9600)}; 
	
	\addplot[draw=black, fill = white, postaction={pattern = crosshatch, pattern color = bcolor8}, every node near coord/.append style={fill = white, opacity = 1}] coordinates {($ 1 $, 36.1500) ($ 2 $, 57.8100) ($ 3 $, 59.1900) ($ 4 $, 59.6000) ($ 5 $, 59.7500)};

\nextgroupplot[
	ymin = 15,
	ymax = 35,
	ylabel = {$ P_\mathrm{tx} $ [dBm]},
	y label style={at={(-0.12,0.5)}, font=\footnotesize,},
	xlabel = {$ N_\mathrm{rx} $},
	x label style={at={(0.5,-0.1)}, font=\footnotesize,}
	]
	
	\addplot[fill = gcolor2] coordinates {($ 1 $, 26.7960) ($ 2 $, 25.4493) ($ 3 $, 25.0416) ($ 4 $, 24.0759) ($ 5 $, 23.6952)}; 
	                   
	\addplot[draw=black, fill = white, postaction={pattern = crosshatch, pattern color = bcolor8}, every node near coord/.append style={fill = white, opacity = 1}] coordinates {($ 1 $, 24.8593) ($ 2 $, 33.0628) ($ 3 $, 33.4899) ($ 4 $, 31.6571) ($ 5 $, 29.3680)}; 
	
	\end{groupplot}
	\end{tikzpicture}
	\vspace{-5mm}
	\caption{Evaluation of the number of decoded packets and transmit power when $ N_\mathrm{rx} $ is varied.}
	\label{f2}
	\vspace{-5mm}
\end{figure}
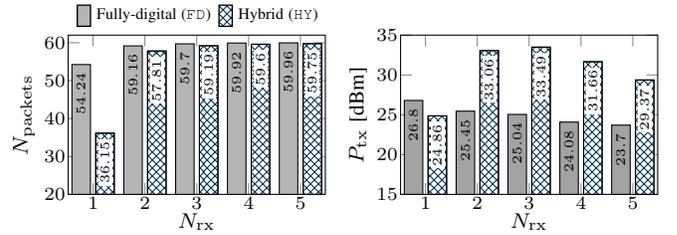

\subsubsection{Impact of $ N_\mathrm{rx} $} The objective of this configuration is to observe the performance improvement of $ N_\mathrm{packets} $ and the importance of the multi-antenna architecture at the receiver. We consider that $ \gamma = 5 $ dB and $ N_\mathrm{iter} = 4 $. Since we vary $ N_\mathrm{rx} $, the number of randomization $ N_{\mathrm{rand}} $ should scale with the dimensionality of $\mathbf{H}_k, \mathbf{F}, \left\lbrace \mathbf{m}_i \right\rbrace^G_{i=1}, \left\lbrace \mathbf{w}_k \right\rbrace^K_{k=1} $. Thus, for this scenario, we select $ N_{\rm{rand}} = 400 + 300 \left( N_\mathrm{tx} + N_\mathrm{rx} - 11 \right) $ with $ N_\mathrm{tx} = 12 $ and $ N_\mathrm{rx} = \left\lbrace 1, 2, 3, 4, 5 \right\rbrace $. For the hybrid precoder, we assume that $ N_{\mathrm{RF}} = 8 $. On the other hand, for the fully-digital version $ N_\mathrm{tx} = N^{\mathrm{RF}}_\mathrm{tx} $. The results in \fref{f2} demonstrate that, with only $ N_\mathrm{rx} = 2 $ antennas at the receiver, it is possible to mitigate the interference and improve $ N_\mathrm{packets} $ considerably. The gain is more noticeable for the hybrid precoder as $ N_\mathrm{packets} $ improves by $ 60\% $. For the fully-digital precoder, there is also a moderate gain of $ 9\% $. Moreover, the average transmit power per successfully received message improves by $ 12.9\% $ and $ 16.8\% $ for the digital and hybrid precoders, respectively. It is evident from this scenario that, at the transmitter side, $ N_\mathrm{packets} $ cannot be further improved when the receivers operate with a single omnidirectional antenna (i.e., $ N_\mathrm{rx} = 1 $), as interference and desired signals are equally amplified. However, when $ N_\mathrm{rx} = 2 $, the receivers can enforce limited selectivity by rejecting the unwanted interference to a certain extent, thereby improving $ N_\mathrm{packets} $. Finally, we observe that $ N_\mathrm{packets} $ for both types of precoders are very similar when $ N_\mathrm{rx} \geq 2 $ although the consumed power differs in $ 6 - 8 $ dBm. 

\subsubsection{Impact of $ N_{\mathrm{rand}} $ and $ N_{\mathrm{iter}} $} 
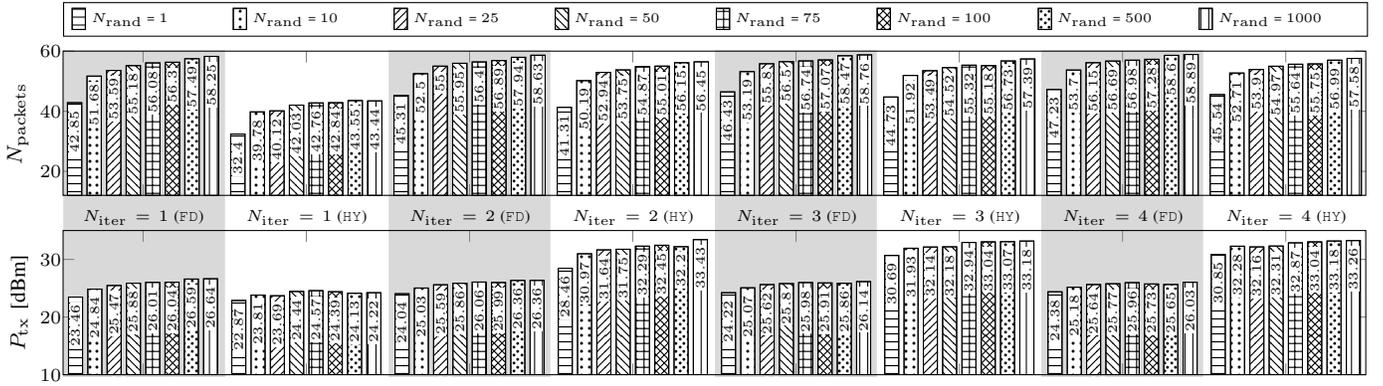
\begin{figure*}[!t]
	
	\centering
	\begin{tikzpicture}[inner sep = 0.2mm]
	\draw [fill=gray, opacity=0.3, draw=gray!0] (0.0,-2.42) rectangle +(2.15,4.35);
	\draw [fill=gray, opacity=0.3, draw=gray!0] (4.325,-2.42) rectangle +(2.15,4.35);
	\draw [fill=gray, opacity=0.3, draw=gray!0] (8.66,-2.42) rectangle +(2.15,4.35);
	\draw [fill=gray, opacity=0.3, draw=gray!0] (13,-2.42) rectangle +(2.15,4.35);
	\begin{axis}
	[
	name = plot1,
	ybar,
	ymin = 12,
	ymax = 60,
	width = 18.9cm,
	height = 3.5cm,
	bar width = 5.4pt,
	tick align = inside,
	x label style={align=center, font=\footnotesize,},
	ylabel = {$ N_\mathrm{packets} $},
	y label style={at={(-0.025,0.5)}, font=\footnotesize,},
	nodes near coords,
	every node near coord/.append style={color = black, rotate = 90, anchor = east, font = \fontsize{2}{2}\selectfont, xshift = -2pt},
	nodes near coords align = {vertical},
	symbolic x coords = {$N_{\rm{iter}} = 1$ (FD), $N_{\rm{iter}} = 1$ (HY), $N_{\rm{iter}}  = 2$ (FD), $N_{\rm{iter}} = 2$ (HY), $N_{\rm{iter}} = 3$ (FD), $N_{\rm{iter}} = 3$ (HY), $N_{\rm{iter}} = 4$ (FD), $N_{\rm{iter}} = 4$ (HY)},
	x tick label style = {text width = 2cm, align = center, font = \fontsize{6}{7}\selectfont, color = black, , yshift=-5pt},
	y tick label style = {font = \fontsize{7}{8}\selectfont},
	xticklabels = {$N_{\rm{iter}} = 1$ (\texttt{FD}), $N_{\rm{iter}}  = 2$ (\texttt{FD}), $N_{\rm{iter}} = 3$ (\texttt{FD}), $N_{\rm{iter}} = 4$ (\texttt{FD}), $N_{\rm{iter}} = 1$ (\texttt{HY}), $N_{\rm{iter}} = 2$ (\texttt{HY}), $N_{\rm{iter}} = 3$ (\texttt{HY}), $N_{\rm{iter}} = 4$ (\texttt{HY})},
	xtick = data,
	enlarge y limits = {value = 0.00, upper},
	enlarge x limits = 0.07,
	legend columns = 8,
	legend pos = north east,
	legend style={at={(-0.00,1.1)}, anchor=south west, font=\fontsize{5}{4}\selectfont, text width = 1.81cm,text height=0.02cm,text depth=.ex, fill = none},
	]

	\addplot[draw=black, fill = white, postaction={pattern = horizontal lines, pattern color = black}, every node near coord/.append style={fill = white, opacity = 1}] coordinates {($N_{\rm{iter}} = 1$ (FD), 42.8500) ($N_{\rm{iter}}  = 2$ (FD), 45.3100) ($N_{\rm{iter}} = 3$ (FD), 46.4300) ($N_{\rm{iter}} = 4$ (FD), 47.2300) ($N_{\rm{iter}} = 1$ (HY), 32.4000) ($N_{\rm{iter}} = 2$ (HY), 41.3100) ($N_{\rm{iter}} = 3$ (HY), 44.7300) ($N_{\rm{iter}} = 4$ (HY), 45.5400)}; 
	\addlegendentry{ $N_{\rm{rand}}$ = $1$ }
	
	\addplot[draw=black, fill = white, postaction={pattern = dots, pattern color = black}, every node near coord/.append style={fill = white, opacity = 1}] coordinates {($N_{\rm{iter}} = 1$ (FD), 51.6800) ($N_{\rm{iter}}  = 2$ (FD), 52.5000) ($N_{\rm{iter}} = 3$ (FD), 53.1900) ($N_{\rm{iter}} = 4$ (FD), 53.7000) ($N_{\rm{iter}} = 1$ (HY), 39.7800) ($N_{\rm{iter}} = 2$ (HY), 50.1900) ($N_{\rm{iter}} = 3$ (HY), 51.9200) ($N_{\rm{iter}} = 4$ (HY), 52.7100)}; 
	\addlegendentry{$ N_{\rm{rand}}$ = $10$}
	
	\addplot[draw=black, fill = white, postaction={pattern = north east lines, pattern color = black}, every node near coord/.append style={fill = white, opacity = 1}] coordinates {($N_{\rm{iter}} = 1$ (FD), 53.5900) ($N_{\rm{iter}}  = 2$ (FD), 55.0000) ($N_{\rm{iter}} = 3$ (FD), 55.8000) ($N_{\rm{iter}} = 4$ (FD), 56.1500) ($N_{\rm{iter}} = 1$ (HY), 40.1200) ($N_{\rm{iter}} = 2$ (HY), 52.9400) ($N_{\rm{iter}} = 3$ (HY), 53.4900) ($N_{\rm{iter}} = 4$ (HY), 53.9000)}; \addlegendentry{$ N_{\rm{rand}}$ = $25$ }
	
	\addplot[draw=black, fill = white, postaction={pattern = north west lines, pattern color = black}, every node near coord/.append style={fill = white, opacity = 1}] coordinates {($N_{\rm{iter}} = 1$ (FD), 55.1800) ($N_{\rm{iter}}  = 2$ (FD), 55.9500) ($N_{\rm{iter}} = 3$ (FD), 56.5000) ($N_{\rm{iter}} = 4$ (FD), 56.6900) ($N_{\rm{iter}} = 1$ (HY), 42.0300) ($N_{\rm{iter}} = 2$ (HY), 53.7500) ($N_{\rm{iter}} = 3$ (HY), 54.5200) ($N_{\rm{iter}} = 4$ (HY), 54.9700)}; \addlegendentry{$ N_{\rm{rand}}$ = $50$ }
	
	\addplot[draw=black, fill = white, postaction={pattern = grid, pattern color = black}, every node near coord/.append style={fill = white, opacity = 1}] coordinates {($N_{\rm{iter}} = 1$ (FD), 56.0800) ($N_{\rm{iter}}  = 2$ (FD), 56.4000) ($N_{\rm{iter}} = 3$ (FD), 56.7400) ($N_{\rm{iter}} = 4$ (FD), 56.9800) ($N_{\rm{iter}} = 1$ (HY), 42.7600) ($N_{\rm{iter}} = 2$ (HY), 54.8700) ($N_{\rm{iter}} = 3$ (HY), 55.3200) ($N_{\rm{iter}} = 4$ (HY), 55.6400)}; \addlegendentry{$ N_{\rm{rand}}$ = $75$ }
	
	\addplot[draw=black, fill = white, postaction={pattern = crosshatch, pattern color = black}, every node near coord/.append style={fill = white, opacity = 1}] coordinates {($N_{\rm{iter}} = 1$ (FD), 56.3000) ($N_{\rm{iter}}  = 2$ (FD), 56.8900) ($N_{\rm{iter}} = 3$ (FD), 57.0700) ($N_{\rm{iter}} = 4$ (FD), 57.2800) ($N_{\rm{iter}} = 1$ (HY), 42.8400) ($N_{\rm{iter}} = 2$ (HY), 55.0100) ($N_{\rm{iter}} = 3$ (HY), 55.1800) ($N_{\rm{iter}} = 4$ (HY), 55.7500)}; \addlegendentry{$ N_{\rm{rand}}$ = $100$ }
	
	\addplot[draw=black, fill = white, postaction={pattern = crosshatch dots, pattern color = black}, every node near coord/.append style={fill = white, opacity = 1}] coordinates {($N_{\rm{iter}} = 1$ (FD), 57.4900) ($N_{\rm{iter}}  = 2$ (FD), 57.9400) ($N_{\rm{iter}} = 3$ (FD), 58.4700) ($N_{\rm{iter}} = 4$ (FD), 58.6000) ($N_{\rm{iter}} = 1$ (HY), 43.5500) ($N_{\rm{iter}} = 2$ (HY), 56.1500) ($N_{\rm{iter}} = 3$ (HY), 56.7300) ($N_{\rm{iter}} = 4$ (HY), 56.9900)}; \addlegendentry{$ N_{\rm{rand}}$ = $500$ }
	
	\addplot[draw=black, fill = white, postaction={pattern = vertical lines, pattern color = black}, every node near coord/.append style={fill = white, opacity = 1}] coordinates {($N_{\rm{iter}} = 1$ (FD), 58.2500) ($N_{\rm{iter}}  = 2$ (FD), 58.6300) ($N_{\rm{iter}} = 3$ (FD), 58.7600) ($N_{\rm{iter}} = 4$ (FD), 58.8900) ($N_{\rm{iter}} = 1$ (HY), 43.4400) ($N_{\rm{iter}} = 2$ (HY), 56.4500) ($N_{\rm{iter}} = 3$ (HY), 57.3900) ($N_{\rm{iter}} = 4$ (HY), 57.5800)}; \addlegendentry{$ N_{\rm{rand}}$ = $1000$ }

	\end{axis}
	
	\begin{axis}
	[
	name=plot2,
	at=(plot1.below south west), anchor=above north west,
	ybar,
	ymin = 10,
	ymax = 35,
	width = 18.9cm,
	height = 3.5cm,
	bar width = 5.3pt,
	tick align = inside,
	x label style={align=center, font=\footnotesize,},
	ylabel = {$ P_\mathrm{tx} $ [dBm]},
	y label style={at={(-0.025,0.5)}, font=\footnotesize,},
	nodes near coords,
	every node near coord/.append style={color = black, rotate = 90, anchor = east, font = \fontsize{2}{2}\selectfont, xshift = -2pt},
	nodes near coords align = {vertical},
	symbolic x coords = {$ N = 1 $, $ N = 2 $, $ N = 3 $, $ N = 4 $, $ N = 5 $, $ N = 6 $, $ N = 7 $, $ N = 8 $},
	x tick label style = {text width = 2cm, align = center, font = \fontsize{7}{8}\selectfont, color=white},
	y tick label style = {font = \fontsize{7}{8}\selectfont},
	xtick = data,
	enlarge y limits = {value = 0.00, upper},
	enlarge x limits = 0.07,
	legend columns = 8,
	legend pos = north east,
	legend style={at={(0.14,0.62)},anchor=south west, font=\fontsize{5}{4}\selectfont, text width = 0.9cm,text height = 0.02cm, text depth=.ex, fill = none}
	]

	
	\addplot[draw=black, fill = white, postaction={pattern = horizontal lines, pattern color = black}, every node near coord/.append style={fill = white, opacity = 1}] coordinates {($ N = 1 $, 23.4638) ($ N = 3 $, 24.0424) ($ N = 5 $, 24.2178) ($ N = 7 $, 24.3756) ($ N = 2 $, 22.8747) ($ N = 4 $, 28.4637) ($ N = 6 $, 30.6880) ($ N = 8 $, 30.8508)};
	
	\addplot[draw=black, fill = white, postaction={pattern = dots, pattern color = black}, every node near coord/.append style={fill = white, opacity = 1}] coordinates {($ N = 1 $, 24.8436) ($ N = 3 $, 25.0288) ($ N = 5 $, 25.0684) ($ N = 7 $, 25.1765) ($ N = 2 $, 23.8102) ($ N = 4 $, 30.9732) ($ N = 6 $, 31.9274) ($ N = 8 $, 32.2787)};
	
	\addplot[draw=black, fill = white, postaction={pattern = north east lines, pattern color = black}, every node near coord/.append style={fill = white, opacity = 1}] coordinates {($ N = 1 $, 25.4729) ($ N = 3 $, 25.5918) ($ N = 5 $, 25.6249) ($ N = 7 $, 25.6360) ($ N = 2 $, 23.6911) ($ N = 4 $, 31.6444) ($ N = 6 $, 32.1387) ($ N = 8 $, 32.1603)};
	
	\addplot[draw=black, fill = white, postaction={pattern = north west lines, pattern color = black}, every node near coord/.append style={fill = white, opacity = 1}] coordinates {($ N = 1 $, 25.8763) ($ N = 3 $, 25.8602) ($ N = 5 $, 25.7993) ($ N = 7 $, 25.7714) ($ N = 2 $, 24.4378) ($ N = 4 $, 31.7512) ($ N = 6 $, 32.1779) ($ N = 8 $, 32.3097)};
	         
	\addplot[draw=black, fill = white, postaction={pattern = grid, pattern color = black}, every node near coord/.append style={fill = white, opacity = 1}] coordinates {($ N = 1 $, 26.0096) ($ N = 3 $, 26.0616) ($ N = 5 $, 25.9789) ($ N = 7 $, 25.9607) ($ N = 2 $, 24.5749) ($ N = 4 $, 32.2890) ($ N = 6 $, 32.9367) ($ N = 8 $, 32.8691)};
	
	\addplot[draw=black, fill = white, postaction={pattern = crosshatch, pattern color = black}, every node near coord/.append style={fill = white, opacity = 1}] coordinates {($ N = 1 $, 26.0429) ($ N = 3 $, 25.9911) ($ N = 5 $, 25.9140) ($ N = 7 $, 25.7316) ($ N = 2 $, 24.3887) ($ N = 4 $, 32.4472) ($ N = 6 $, 33.0412) ($ N = 8 $, 33.0373)};
	
	\addplot[draw=black, fill = white, postaction={pattern = crosshatch dots, pattern color = black}, every node near coord/.append style={fill = white, opacity = 1}] coordinates {($ N = 1 $, 26.5852) ($ N = 3 $, 26.3563) ($ N = 5 $, 25.8593) ($ N = 7 $, 25.6527) ($ N = 2 $, 24.1289) ($ N = 4 $, 32.2012) ($ N = 6 $, 33.0748) ($ N = 8 $, 33.1842)};
	
	\addplot[draw=black, fill = white, postaction={pattern = vertical lines, pattern color = black}, every node near coord/.append style={fill = white, opacity = 1}] coordinates {($ N = 1 $, 26.6354) ($ N = 3 $, 26.3584) ($ N = 5 $, 26.1436) ($ N = 7 $, 26.0347) ($ N = 2 $, 24.2198) ($ N = 4 $, 33.4269) ($ N = 6 $, 33.1754) ($ N = 8 $, 33.2625)};

	\end{axis}
	\end{tikzpicture}
	\vspace{-5mm}
	\caption{Evaluation of the number of decoded packets and transmit power for $ N_\mathrm{tx} = 12 $ when $ N_{\mathrm{iter}} $ and $ N_{\mathrm{rand}} $ are varied.}
	\label{f3}
	\vspace{-3mm}
\end{figure*}

The objective of this setting is to analyze the performance sensitivity of the fully-digital and hybrid precoders to the selection of $ N_{\mathrm{rand}} $ and $ N_{\mathrm{iter}} $. To this purpose, we consider $ N_\mathrm{rx} = 2 $ and $ \gamma = 5 $ dB. For the fully-digital precoder, we assume that $ N_\mathrm{tx} = N^{\mathrm{RF}}_\mathrm{tx} = 12 $. On the other hand, the hybrid precoder is endowed with $ N^{\mathrm{RF}}_\mathrm{tx} = 8 $ and $ N_\mathrm{tx} = 12 $. We evaluate the performance variation when $ N_{\mathrm{iter}} = \left\lbrace 1, 2, 3, 4, 5 \right\rbrace $ and $ N_{\mathrm{rand}} = \left\lbrace 1, 10, 25, 50, 75, 100, 500, 1000  \right\rbrace $. The results in terms of $ N_\mathrm{packets} $ and $ P_\mathrm{tx} $ for both types of precoder are shown in Fig. \ref{f3}. We observe that in the fully-digital precoder case, $ N_{\mathrm{rand}} $ is more influential than $ N_{\mathrm{iter}} $ since $ N_\mathrm{packets} $ improves noticeably when $ N_{\mathrm{rand}} $ is augmented, whereas a small improvement can be observed between the cases $ N_\mathrm{iter} = 1 $ and $ N_\mathrm{iter} = 2 $. Conversely, for the hybrid precoder, $ N_{\mathrm{iter}} $ promotes performance gap reduction with respect to the fully-digital implementation. The fully-digital precoder converges faster since only $ \left\lbrace \mathbf{m}_i \right\rbrace^G_{i=1} $ and $ \left\lbrace \mathbf{w}_k \right\rbrace^K_{k=1} $ are optimized. In the hybrid precoder case, we need to design $ \mathbf{F} $, $ \left\lbrace \mathbf{m}_i \right\rbrace^G_{i=1} $ and $ \left\lbrace \mathbf{w}_k \right\rbrace^K_{k=1} $ with even more limiting constraints (finite-resolution constant-modulus phase shifts), thus requiring more iterations to obtain an stable solution.

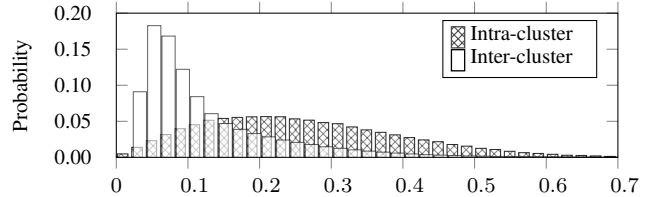
\begin{figure}[!tb]
	\centering
	\begin{tikzpicture}
		\begin{axis}
			[
			ybar,
			ymin = 0.00,
			ymax = 0.2,
			xmax = 0.7,
			xmin = 0,
			width = 8.25cm,
			height = 3.5cm,
			x label style={align=center, font=\footnotesize,},
			ylabel = {Probability},
			y label style={at={(-0.15,0.5)}, font=\footnotesize,},
			x tick label style = {text width = 0.2cm, align = center, font = \fontsize{8}{9}\selectfont},
			y tick label style = {text width = 0.9cm, align = center, font = \fontsize{8}{9}\selectfont},
			ytick = {0.00, 0.05, 0.10, 0.15, 0.20},
			yticklabels = {0.00, 0.05, 0.10, 0.15, 0.20},
			legend columns = 1,
			legend pos = north east,
			legend style={at={(0.65,0.55)},anchor=south west, font=\fontsize{8}{9}\selectfont, text 	width=1.5cm,text height=0.02cm,text depth=.ex, fill = none}
			]
			
			\addlegendimage{draw=p7, fill = gcolor5!60, mark=none, line width=0.5pt, pattern = crosshatch, pattern color = p7},
			\addlegendentry{Intra-cluster},
			\addlegendimage{draw=black, fill = white!80, mark=none, line width=0.5pt},
			\addlegendentry{Inter-cluster},
			
			\addplot[ybar, bar width = 4pt, fill = white!60, opacity = 0.8, pattern = crosshatch, pattern color = p7] 
			table[x index = 0, y index = 1] {IntraGroupCorr.txt};
					
			\addplot[ybar, bar width = 5pt, fill = white!80, opacity = 0.6] 
			table[x index = 0, y index = 1] {InterGroupCorr.txt}; 
			
			\end{axis}
	\end{tikzpicture}
	\vspace{-4mm}
	\caption{Channel correlation histogram.}
	\label{f4}
	\vspace{-3mm}
\end{figure}

\begin{figure}[!tb]
	\centering
	\begin{tabular}[t]{cccc}
		\begin{subfigure}{0.3\columnwidth}
			\begin{tikzpicture}
				\begin{polaraxis}
				[
				domain = -90: 90,
				xmin = -90, 
				xmax = 90,
				width = 6.0cm,
				height = 6.0cm,
				xticklabel = $\pgfmathprintnumber{\tick}^\circ$,
				x tick label style = {font = \fontsize{9}{10}\selectfont},
				y tick label style = {font = \fontsize{9}{10}\selectfont},
				xtick = {-80, -60, -40, -20, 0, 20, 40, 60, 80},
				ytick = {0, 0.1, 0.2, 0.3, 0.4},	
				]
				\addplot [no markers, thick, bcolor6] table {PatternTX.txt};
				\end{polaraxis}
				
				\node[below] at (1.4, -0.6) {\footnotesize{Transmitter}};
				
				\draw[fill = gcolor7] (1.0, 0.3) circle (0.2cm) node[text=white] {$ {\scriptsize \text{U}_1} $};
				\draw[fill = gcolor7] (2.0, 1.45) circle (0.2cm) node[text=white] {$ {\scriptsize \text{U}_2} $};
				\draw[fill = gcolor7] (2.0, 3.0) circle (0.2cm) node[text=white] {$ {\scriptsize \text{U}_3} $};
				\draw[fill = gcolor7] (1.0, 4.1) circle (0.2cm) node[text=white] {$ {\scriptsize \text{U}_4} $};
			\end{tikzpicture}
		\end{subfigure}
		&
		\begin{tabular}{c}
			\smallskip
			\begin{subfigure}[t]{0.16\columnwidth}
				\centering
				\begin{tikzpicture}
					\begin{polaraxis}
					[
					domain = 90: 270,
					xmin = 90, 
					xmax = 270,
					ymax = 0.01,
					width = 3.0cm,
					height = 3.0cm,
					xticklabel = $\pgfmathprintnumber{\tick}^\circ$,
					xtick = {135, 180, 225},
					every y tick scale label/.style={at={(1,1.15)}},
					ytick = {0.005, 0.01},	
					yticklabels = {\empty},
					x tick label style = {font = \fontsize{9}{10}\selectfont},
					y tick label style = {font = \fontsize{9}{10}\selectfont},
					]
					\addplot [no markers, thick, bcolor6] table {PatternRX1.txt};
					\end{polaraxis}
					\node[below] at (0.6, -0.25) {\footnotesize{User 1} ($ {\scriptsize \text{U}_1} $)};
				\end{tikzpicture}
			\end{subfigure}\\
			\begin{subfigure}[t]{0.16\columnwidth}
				\centering
				\begin{tikzpicture}
					\begin{polaraxis}
					[
					domain = 90: 270,
					xmin = 90, 
					xmax = 270,
					ymax = 0.01,
					width = 3.0cm,
					height = 3.0cm,
					xticklabel = $\pgfmathprintnumber{\tick}^\circ$,
					xtick = {135, 180, 225},
					every y tick scale label/.style={at={(1,1.15)}},
					ytick = {0.005, 0.01},	
					yticklabels = {\empty},	
					x tick label style = {font = \fontsize{9}{10}\selectfont},
					y tick label style = {font = \fontsize{9}{10}\selectfont},	
					]
					\addplot [no markers, thick, bcolor6] table {PatternRX3.txt};
					\end{polaraxis}
					\node[below] at (0.6, -0.25) {\footnotesize{User 3} ($ {\scriptsize \text{U}_3} $)};
				\end{tikzpicture}
			\end{subfigure}\\
		\end{tabular}
		&
		\begin{tabular}{c}
			\smallskip
			\begin{subfigure}[t]{0.2\columnwidth}
				\centering
				\begin{tikzpicture}
				\begin{polaraxis}
				[
				domain = 90: 270,
				xmin = 90, 
				xmax = 270,
				ymax = 0.01,
				width = 3.0cm,
				height = 3.0cm,
				xticklabel = $\pgfmathprintnumber{\tick}^\circ$,	
				xtick = {135, 180, 225},
				every y tick scale label/.style={at={(1,1.15)}},
				ytick = {0.005, 0.01},	
				yticklabels = {\empty},
				x tick label style = {font = \fontsize{9}{10}\selectfont},
				y tick label style = {font = \fontsize{9}{10}\selectfont},	
				]
				\addplot [no markers, thick, bcolor6] table {PatternRX2.txt};
				\end{polaraxis}
				\node[below] at (0.6, -0.25) {\footnotesize{User 2} ($ {\scriptsize \text{U}_2} $)};
				\end{tikzpicture}
			\end{subfigure}\\
			\begin{subfigure}[t]{0.2\columnwidth}
				\centering
				\begin{tikzpicture}
				\begin{polaraxis}
				[
				domain = 90: 270,
				xmin = 90, 
				xmax = 270,
				ymax = 0.01,
				width = 3.0cm,
				height = 3.0cm,
				xticklabel = $\pgfmathprintnumber{\tick}^\circ$,	
				xtick = {135, 180, 225},
				every y tick scale label/.style={at={(1,1.15)}},
				ytick = {0.005, 0.01},	
				yticklabels = {\empty},	
				x tick label style = {font = \fontsize{9}{10}\selectfont},
				y tick label style = {font = \fontsize{9}{10}\selectfont},
				]
				\addplot [no markers, thick, bcolor6] table {PatternRX4.txt};
				\end{polaraxis}
				\node[below] at (0.6, -0.25) {\footnotesize{User 4} ($ {\scriptsize \text{U}_4} $)};
				\end{tikzpicture}
			\end{subfigure}\\
		\end{tabular}\\
		&
		\begin{subfigure}{0.01\columnwidth}
		\end{subfigure}
	\end{tabular}
	\vspace{-6mm}
	\caption{Radiation patterns}
	\label{f5}
	\vspace{-5mm}
\end{figure}
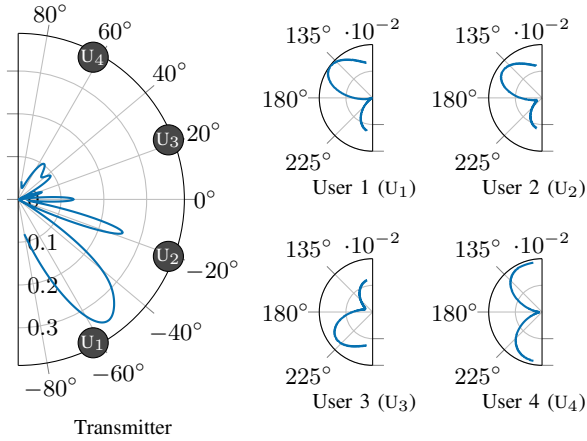

\textbf{Generation of co-channel users:}
In order to gain more understanding on the kind of scenario we are dealing with, we show in Fig. \ref{f4} the histogram of channel correlations for \emph{(i)} users that belong to the same group (intra-cluster) and \emph{(ii)} users that belong to different groups (inter-cluster). In the first case, the average channel correlation is 0.24 whereas in the second case is 0.10. The mean angles of departure $ \left\lbrace \bar{\theta}^{\mathrm{AoD}}_i \right\rbrace^G_{i = 1} $ for the multicast groups are distributed in the range $ \left[ -80 \degree, 80 \degree \right] $ with angular spread $ \sigma_{\mathrm{AoD}} = 30 \degree $. The mean angles of arrival $ \left\lbrace \bar{\theta}^{\mathrm{AoA}}_k \right\rbrace^{K}_{k = 1} $ for each receiver are uniformly distributed in the range $ \left[ -360 \degree, 360 \degree \right] $ with angular spread of $ \sigma_{\mathrm{AoA}} = 60 \degree $. Thus, for a given user $ k' $ that belongs to group $ G_{i'} $, the paths will have angles of departure and arrival in the ranges $ \bar{\theta}^{\mathrm{AoD}}_{i'} \pm \sigma_{\mathrm{AoD}} $ and $ \bar{\theta}^{\mathrm{AoA}}_{i'} \pm \sigma_{\mathrm{AoD}} $, respectively. To shed more light on this aspect, Fig. \ref{f5} shows a particular channel realization when $ N_\mathrm{tx} = 8 $, $ N^{\mathrm{RF}}_\mathrm{tx} = 4 $, $ N_\mathrm{rx} = 2 $, $ K = 4 $, $ G = 4 $, $ \bar{\theta}^{\mathrm{AoD}}_1 = -60\degree $, $ \bar{\theta}^{\mathrm{AoD}}_2 = -20\degree $, $ \bar{\theta}^{\mathrm{AoD}}_3 = 20\degree $, $ \bar{\theta}^{\mathrm{AoD}}_4 = 60\degree $ and $ \sigma_{\mathrm{AoD}} = 5\degree $. Due to existence of multiple paths, the transmit and receive beams are not fully aligned as expected in line-of-sight scenarios. Thus, each of the users orients their receive power in specific directions that are coherent with the most meaningful beams of the transmitter beam-pattern. Also, note that secondary lobes at the receiver have been shaped to minimize amplification of interfering signals.

%% file: discussion.tex
\section{Discussion}
\label{sec:discussion}

\noindent{\textbf{Co-channel users:}} We have considered a very challenging scenario throughout our simulations in order to examine the operational limits of our design. We can observe from Fig. \ref{f4} that intra-cluster and inter-cluster users are not easily separable as a subset of them have similar channel correlations. In our scenarios, this is determined by the selection of $ \left\lbrace \bar{\theta}^{\mathrm{AoD}}_i \right\rbrace^G_{i = 1} $, $ \left\lbrace \bar{\theta}^{\mathrm{AoA}}_k \right\rbrace^K_{k = 1} $, $ \sigma_{\mathrm{AoD}} $ and $ \sigma_{\mathrm{AoA}} $. We notice that highly correlated users (co-located users) belonging to different groups were the most challenging to cater, specifically for the hybrid precoder whose beamforming flexibility is limited.
 
\noindent{\textbf{Hybrid precoder design:}} Different from the majority of works in hybrid precoding (either multi-user or multicast), the proposed design has no knowledge of the optimal fully-digital precoder (as in e.g., \cite{b17}). Thus, our proposed design is not obtained as an approximation of the optimal fully-digital implementation. Without an optimal reference, the design becomes more challenging.

\noindent{\textbf{Initial points:}} We have considered naive initializations for the optimization parameters. We leave for future work the exploitation of AoA/AoD to infer more befitting initializations and thus improve the performance of the scheme. 

\noindent{\textbf{Fully-digital precoder design:}} The fully-digital implementation is obtained by assigning $ \mathbf{F} = \mathbf{I} $ and then optimizing alternately over $ \left\lbrace \mathbf{m}_i \right\rbrace^G_{i = 1} $ and $ \left\lbrace \mathbf{w}_k \right\rbrace^K_{k = 1} $.

\noindent{\textbf{Algorithm convergence:}} There is no theoretical evidence supporting the convergence of Algorithm \ref{a1}, essentially due to the non-convexity of the problem. However, the proposed scheme exhibits an stable behavior for both digital and hybrid precoders since the solutions do not vary significantly as $ N_\mathrm{iter} $ and $ N_\mathrm{rand} $ increase beyond a certain limit.

\noindent{\textbf{Computational complexity:}} Neglecting the complexity owing to randomization and obviating the insignificant complexity increase due to the inclusion of slack parameters, the computational complexity of the proposed scheme when $ N_\mathrm{iter} = 1 $ is $ \mathcal{O} \left( \left( N^{\mathrm{RF}}_\mathrm{tx} N_\mathrm{tx} \right)^6 + K \left( N^{\mathrm{RF}}_\mathrm{tx} N_\mathrm{tx} \right)^2 \right) + \mathcal{O} \left( G^3 \left( N^{\mathrm{RF}}_\mathrm{tx} \right)^6 + K G \left( N^{\mathrm{RF}}_\mathrm{tx} \right)^2 \right) + \mathcal{O} \left( K \left( N_\mathrm{rx} \right)^6 + K \left( N_\mathrm{rx} \right)^4 \right) $.

%% file: conclusion.tex
\section{Conclusion}
\label{sec:conclusion}
In this paper, we investigated the optimization of multi-group multicast hybrid precoders in mmWave systems. Our proposed solution is based on the alternating optimization, semidefinite relaxation and Cholesky matrix factorization, where the digital precoder, analog phase shifts, and receive combiners are optimized sequentially in an iterative manner. Furthermore, our formulation allows the employment of an arbitrary number of phase shifts. It was corroborated through extensive simulations that the hybrid precoder can indeed attain similar performance as its fully-digital counterpart, even in very challenging scenarios with high inter-cluster user correlation. In addition, we demonstrate that having receivers with two antennas suffices to improve the number of decoded packets. Thus, our proposed design achieves up to $60\%$ gain.

%% file: acknowledgment.tex
\section{Acknowledgment}
\label{sec:acknowledgment}
This work has been funded by the LOEWE initiative (Hessen, Germany) within the German Research Foundation (DFG) in the Collaborative Research Center (CRC) 1053 - MAKI.